\documentclass[twocolumn]{aastex62}
\pdfoutput=1 
\usepackage{amsmath,amstext}
\usepackage[T1]{fontenc}
\usepackage{apjfonts} 
\usepackage[figure,figure*]{hypcap}

\newcommand{\iso}[2]{\ensuremath{{}^{#2}\mathrm{#1}}}
\newcommand{\srad}{\ensuremath{S_\mathrm{rad}}}
\newcommand{\feti}{\ensuremath{\mathrm{Fe}/\iso{Ti}{44}}}

\newcommand{\sci}[2]{\ensuremath{{#1} \times 10^{#2}}}
\newcommand{\msun}{\ensuremath{\mathrm{M}_\odot}}
\newcommand{\rsun}{\ensuremath{\mathrm{R}_\odot}}

\shorttitle{Ti and Fe in Cas A}
\shortauthors{Vance et al.}

\begin{document}

\title{Titanium and Iron in the Cassiopeia A Supernova Remnant}

\author{Gregory S. Vance}
\affiliation{School of Earth and Space Exploration, Arizona State University, P.O. Box 1404, Tempe, AZ 85287-1404}

\author{Patrick A. Young}
\affiliation{School of Earth and Space Exploration, Arizona State University, P.O. Box 1404, Tempe, AZ 85287-1404}

\author{Christopher L. Fryer}
\affiliation{Center for Theoretical Astrophysics, Los Alamos National Laboratory, MS D409, Los Alamos, NM 87544}

\author{Carola I. Ellinger}
\affiliation{Center for Theoretical Astrophysics, Los Alamos National Laboratory, MS D409, Los Alamos, NM 87544}

\begin{abstract}
Mixing above the proto-neutron star is believed to play an important role in the supernova engine, and this mixing results in a supernova explosion with asymmetries.  Elements produced in the innermost ejecta, e.g., ${}^{56}$Ni and ${}^{44}$Ti, provide a clean probe of this engine.  The production of ${}^{44}$Ti is particularly sensitive to the exact production pathway and, by understanding the available pathways, we can use ${}^{44}$Ti to probe the supernova engine.  Using thermodynamic trajectories from a three-dimensional supernova explosion model, we review the production of these elements and the structures expected to form under the ``convective-engine'' paradigm behind supernovae.  We compare our results to recent X-ray and $\gamma$-ray observations of the Cassiopeia A supernova remnant.
\end{abstract}

\keywords{Cassiopeia A --- nucleosynthesis --- supernovae}

\section{Introduction}
\label{sec:intro}

Guided by the observed mixing of \iso{Ni}{56} in SN~1987A, astronomers began to develop a model for the core-collapse supernova (CCSN) engine where the efficiency of potential energy release in the collapse is increased through hydrodynamic instabilities above the proto-neutron star \citep{1992ApJ...387..294H}.  The first successful explosion produced by modeling the collapse and engine of a massive star in multiple dimensions demonstrated that this physics was key to the explosion process \citep{1994ApJ...435..339H}.  Twenty-five years later, although the nature of the hydrodynamic instabilities remains a matter of debate, this engine has now become the standard supernova engine \citep[e.g.,][]{fryer07,takiwaki14,melson15,burrows18}.  The basic model argues that after the collapse of a massive star, the core reaches nuclear densities and bounces, driving a shock that soon stalls.  The region between the dense proto-neutron star and the stalled shock is susceptible to a number of convective instabilities.  Convection distributes the energy from the proto-neutron star's surface outwards to the edge of the stalled shock, and it also reduces the mass at the stalled shock by transporting material that was piling up at the shock inwards to the proto-neutron star.  Both of these factors increase the probability of a successful explosion occurring.

This supernova-engine paradigm has undergone a continuous series of verification and validation tests (including a broad range of code-comparison studies such as those listed in the previous paragraph) comparing model predictions to observations.  For example, driven by the need for explosion asymmetries in models of SN~1987A, this engine model argued for asymmetries that could develop at low modes, possibly producing the ``kicks'' seen in pulsars \citep{1995PhR...256..117H,2006A&A...457..963S}.  By noting that the explosion energy for this mechanism is set to the energy stored in the convective region prior to the launch of the shock, engine theorists were able to explain the fact that although the collapse releases $\sim 10^{53}\ \mathrm{erg}$, typical explosion energies are $\sim 10^{51}\ \mathrm{erg}$ \citep{1999ApJ...522..413F}.  At a time when supernova observations predicted that only very massive stars would explode \citep{hamuy02}, this engine argued that only stars with masses (neglecting mass loss) below $\sim 23\ \msun$ would explode \citep{1999ApJ...522..413F}\footnote{Not surprisingly, mass loss can alter this effect and allow more massive stars to explode.}.  Likewise, in a time when the remnant mass distribution was believed to be a set of delta functions \citep{thorsett99}, this model predicted a range of neutron star and black hole remnant masses \citep{fryer01}.  Both of these predictions were ultimately confirmed by later observations \citep{2012ARNPS..62..485L}.

With the acceptance of this paradigm, observations could then be turned to better understand the details of the model.  For example, although the convective engine does not predict a mass gap in the mass distribution of compact remnants, the existence of a mass gap can place constraints on the uncertainties in the engine \citep{fryer12}.  The nucleosynthetic yields also place strong constraints on the supernova engine.  Unfortunately, most elements can only be observed if they are excited by the reverse shock, and any abundance study must incorporate the uncertainties in estimating the distributions and masses of the different elements in the ejecta.  The reverse shock is produced as the supernova shock decelerates in the circumstellar medium, and it is often difficult to distinguish asymmetries in the explosion from asymmetries in the circumstellar medium in these observations.  The exception to these limitations is the measurement of \iso{Ti}{44}.  The decay half-life of \iso{Ti}{44} is $\sim 60\ \mathrm{yr}$, and hence is ideally suited for studies of 100--1000~yr old remnants.  Photons emitted in the radioactive decay of \iso{Ti}{44} and its daughter products are a direct measurement of the \iso{Ti}{44} yield, unaffected by uncertain shock dynamics and asymmetries in the circumstellar medium.  In addition, \iso{Ti}{44} is produced in the innermost ejecta, providing a direct probe of the central core-collapse engine.  NuSTAR observations of the \iso{Ti}{44} distribution in the Cassiopeia A supernova remnant provided the first glimpse of the asymmetries in the supernova engine, as well as a direct confirmation of the low-mode asymmetries predicted by the convective supernova engine \citep{grefenstette14,grefenstette17}.

Discussion of \iso{Ti}{44} and \iso{Ni}{56} production in the inner ejecta of supernovae dates back just as far as investigations of the convective supernova engine \citep[e.g.][]{Theilemann90,Hoffman95,Diehl98}. The two topics are nearly inseparable given the importance of \iso{Ti}{44} and Fe as observational probes of deep interior supernova conditions. \citet{Thielemann96} pointed out that the production of these isotopes depends on radiation entropy and on the strength of the $\alpha$-rich freeze-out. \citep{Hoffman99} discussed equilibrium features and sensitivity to reaction rates. More recently, \citet{Magkotsios10} presented a detailed grid analysis of \iso{Ti}{44} and \iso{Ni}{56} production in CCSNe. The bulk of their analysis studied simple trajectories fitting power laws and exponential decays. A pair of two-dimensional explosions were briefly examined but not analyzed in detail.

Cassiopeia~A (Cas~A) is one of the best-studied supernova remnants, with a broad set of constraints on the progenitor at the time of collapse as well as the explosion energetics and asymmetries \citep{2006ApJ...640..891Y}.  The existence of nitrogen knots in the supernova remnant \citep{fesen01} argues that the stellar material must have undergone CNO processing and the helium envelope should have been exposed \citep{arnett96}.  This argues that the explosion was a type-Ib/IIb supernova, a fact confirmed by spectral and light-curve observations of the light-echo \citep{krause08}.  The ejecta mass is more difficult to predict, and is based both on kinematic properties and emission measures across a wide range of wavelengths.  These models predict ejecta masses between 2 and 4~\msun.  If the total mass of the star at explosion was the remnant mass ($\sim 1.2$--$2\ \msun$) plus the ejecta mass, then the progenitor helium star had a mass of $\sim 3.5$--$6\ \msun$.  This corresponds to an initial progenitor mass of roughly $13$--$23\ \msun$, with the majority of the hydrogen envelope being stripped off by a binary interaction.  These constraints also place limits on the explosion energy \citep{chevalier03}.

Current studies of the nucleosynthetic yields of Cas~A have only scratched the surface of what we might learn from this data. For example, \citet{Wongwathanarat17} published yields for a model resembling Cas~A, but did not include a detailed description of thermodynamic trajectories.  They also assumed a single value of $Y_e$ based on the progenitor composition outside a cutoff radius in their preferred model, which was chosen to better match the Cas~A remnant.  Neutrino processing may affect the final yields, but assuming a single value of $Y_e$ is not realistic \citep{2018IJMPD..2750116S,2019MNRAS.488L.114F}.
\citet{Couch15} simulated the last few minutes of Si-shell burning in 3D before modeling the collapse and explosion, which they also ran in 3D with a single octant and a network of 21 isotopes.  Their star was evolved to Si buring using MESA, which suffers from incorrect treatment of mixing \citep{321d17, am11, amy09}.  \citet{Harris17} ran an explosion engine in 2D out to $t \sim 1\ \mathrm{s}$. They use the \citet{Woosley07} progenitor models evolved using KEPLER, which also lacks an accurate treatment of convection. \citet{Eichler18} ran two simulations of different masses and examined how the varying $Y_e$ values changed the yields of heaver nuclei well beyond the iron group, but their 2D models suffer from dimensionality problems with convection in the explosion.

In this project, we compare the spatial and velocity distributions of the \iso{Ti}{44} in an explosion based on the convective engine to the observed distributions from NuSTAR and XMM. We produce a bimodal distribution in the \feti{} ratio, which we argue could be a common feature of supernova remnants. In addition, we find that deceleration of the shock can re-heat material significantly enough to alter thermodynamic trajectories and the resulting yields. The methods used in these calculations are described in section~\ref{sec:methods}, with the \iso{Ti}{44} distributions presented in section~\ref{sec:tini}.  The \iso{Ti}{44} yields relative to iron-peak elements (also synthesized in the innermost ejecta) provide constraining probes of the strength of the engine, but this requires understanding the detailed thermodynamic trajectories that are produced in these explosions, which are also discussed in section~\ref{sec:tini}.  We conclude with a brief discussion of the other yields of Cas~A.

\section{Methods}
\label{sec:methods}

\subsection{Stellar Models Using Tycho}

The progenitor star was simulated using the stellar evolution code Tycho \citep{yoar05}.  Tycho is a one-dimensional stellar evolution code with a hydrodynamic formulation of the stellar evolution equations.  It uses OPAL and revised low temperature opacities \citep{ir96,alfer94,rn02,f05, sbfa09}, a combined OPAL and Timmes equation of state (HELMHOLTZ) \citep{ta99,rn02}, gravitational settling (diffusion) \citep{tbl94}, general relativistic gravity, time lapse, curvature, automatic rezoning, and an adaptable nuclear reaction network with a sparse solver. A 177-element network terminating at \iso{Ge}{74} is used throughout the evolution.  The network uses the latest REACLIB rates \citep{rat00,ang99,iliadis01,wie06}, weak rates from \citet{lmp00}, and screening from \citet{grab73}. Neutrino cooling from plasma processes and the Urca process is included. 

Mass loss uses a choice of updated versions of the prescriptions of \citep{Kudritzki89} or prescriptions based on \citet{vink01,mokiem07} for OB mass loss, \citet{Bloecker95} for red supergiant mass loss, and \citet{LamersNugis02} for WR phases.

Tycho incorporates a description of turbulent convection projected down to 1D secularly evolving average behavior \citep{321d17}, which is based on three-dimensional, well-resolved simulations of convection sandwiched between stable layers, which were analyzed in detail using a Reynolds decomposition into average and fluctuating quantities \citep{ma07,amy09,amy10,am11,321d17}. Unlike mixing-length theory, it has no free convective parameters to adjust. The inclusion of these processes, which approximate the integrated effect of dynamic stability criteria for convection, entrainment at convective boundaries, and wave-driven mixing, results in significantly larger turbulently mixed regions. Therefore the extent of material processed by particular core or shell burning phases is higher, and the stellar core at a given stage is more massive and denser.

\begin{figure}
\epsscale{1.15}
\plotone{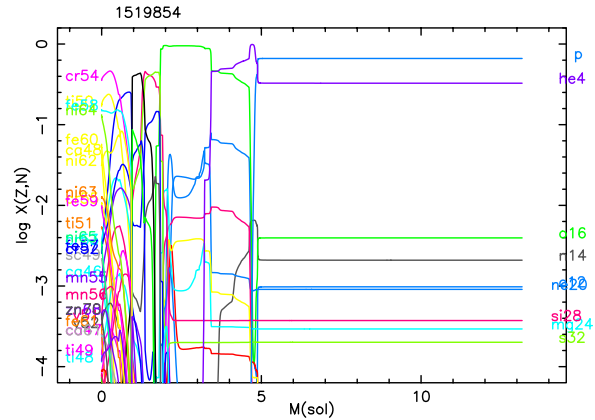}
\caption{Mass fractions for isotopes plotted against mass coordinate (enclosed mass) for the final progenitor star model.}
\label{15-final-Y}
\end{figure}

The progenitor used here is a non-rotating $15\ \msun$ star with solar composition from \citet{lod10}. The mass of the final model used as the initial condition for the supernova calculation was $13.15\ \msun$. 
This model does not include stripping by a binary companion. The retention of a hydrogen envelope would affect the late time behavior of the explosion significantly \citep[][i.e.]{ellinger13}, but the production of \iso{Ti}{44} and \iso{Ni}{56} depends only on the structure of the deep interior of the star. Relatively early mass loss would change the structure of the progenitor star's core, but the best candidate for the progenitor of Cas~A is a binary with late-time envelope ejection \citep{2006ApJ...640..891Y}. Comparing the progenitor model from this calculation with a stripped model of similar mass, the radial profiles inside the H envelope revealed the density and temperature differing by less than a factor of 10. We note that massive star models in the literature suffer from incorrect mixing physics \citep{ma07,amy09,amy10,am11,321d17}, which leads to significantly greater discrepancies. At minimum, this difference makes our model no less relevant than models presented in other works. Evolution was terminated when negative velocities existed throughout the core and the central temperature rose to $T > \sci{5}{9}\ \mathrm{K}$. Figure~\ref{15-final-Y} shows the mass fraction of isotopes plotted against mass coordinate (enclosed mass) for the final model.

\subsection{Collapse and Explosion Models}

To model the stellar collapse and ensuing explosion, we use a one-dimensional Lagrangian code to follow the collapse through core bounce.  This code includes three-flavor neutrino transport using a flux-limited diffusion calculation and a coupled set of equations of state to model the wide range of densities in the collapse phase \citep[see][for details]{1994ApJ...435..339H, 1999ApJ...522..413F}.  It includes a 14-element nuclear network \citep{1989ApJ...342..986B} to follow the energy generation.   

The shock is then revived by adding an energy injection following the parameterized method of \cite{2018ApJ...856...63F}.  In this model, roughly $\sci{5}{51}\ \mathrm{erg}$ was deposited into the inner $0.02\ \msun$ in the first $150\ \mathrm{ms}$.  Some of this energy is lost through neutrino emission and the total explosion energy at late times for this model is $\sci{1.5}{51}\ \mathrm{erg}$.  This explosion is then mapped into our three-dimensional calculations, using 1~million SPH particles. The mapping took place when the supernova shock had moved out of the iron core and propagated into the Si-S rich shell at $t < 1\ \mathrm{s}$. We note that our 1D methods employed for modeling the collapse, core bounce, and initial explosion do not capture the full physics of the central engine \citep[for a discussion, see][]{2018ApJ...856...63F}, and this is a source of uncertainty in our yields calculations. The details of the engine change the shock trajectories, and neutrino chemistry can change $Y_e$ values \citep{2018IJMPD..2750116S,2019MNRAS.488L.114F}. The nature of the shock affects mostly the yields after the shock falls below NSE (before it falls out of NSE, the yields are set by the equilibrium values, not the time-dependent evolution). Our model captures one instance of the range of asymmetric trajectories, and it should be noted that no model at this time is sufficiently accurate to dictate exactly the properties of the asymmetries \citep{2016ARNPS..66..341J}. In addition, any model that does not include convection-driven asymmetries from the progenitor star cannot properly capture the asymmetries \citep{Arnett312D2015}. The 3D explosion model used here also displays stochastic asymmetries, implying that any manner of convective asymmetry could generate similar results. If this behavior is universal, it could have important implications. These points taken together indicate that nucleosynthetic patterns arising from convection-like behavior are robust, regardless of the driver. As discussed below, this increases the utility of NSE nucleosynthesis, particularly of \iso{Ti}{44} and \iso{Ni}{56}, as diagnostics of the conditions in the progenitor star.

The latter point, the $Y_e$ values, could alter our results as well \citep[see, e.g.,][]{Magkotsios10}. Although more detailed models have addressed neutrino interactions and the evolution of $Y_e$, the neutrino physics is not yet sufficiently accurate to model this correctly \citep{2018IJMPD..2750116S}. In light of this, any nucelosynthetic calculation under these conditions is subject to uncertainty. In order to assess the effect of these uncertainties on the results presented here, we carried out a series of nucleosynthesis calculations using thermodynamic trajectories from the explosion model with a range of values for $Y_e$. The vast majority of material that reached sufficiently high temperatures for production of \iso{Ti}{44} and \iso{Ni}{56} saw yield changes of no more than 3--4\% for $Y_e$ values from 0.495 to 0.499. This is a larger range of $Y_e$ than we would expect from the material ejected in this particular explosion, so these results provide a conservative bound on our uncertainties. As discussed later in Section~\ref{sec:tini}, many of our more important results are general enough to be robust to small changes in the nucleosynthetic conditions, so we expect these uncertainties to not qualitatively alter our conclusions.

The
3D simulation used the SNSPH smoothed particle hydrodynamics (SPH) code \citep{fryer06} to follow the long-term evolution of the supernova explosion and remnant.  This code has been extensively used to follow the ejecta of supernovae \citep{hungerford03,hungerford05,young09,ellinger12,ellinger13,wong14}, taking advantage of the adaptive time steps and variable particle scale lengths in the method.  A central gravity source with absorbing boundary was included to simulate a compact central object (CCO) with an initial mass of $1.35\ \msun$ and radius of $10^{-4}\ \rsun$. Total mass, linear momentum, and angular momentum accreted onto the central object were tracked.

The SNSPH code makes use of a limited nuclear reaction network of 20 isotopes to expedite the energy calculations for the hydrodynamics.  The network terminates at \iso{Ni}{56} and neutron excess is directed to \iso{Fe}{54}.  The network runs in parallel to the hydrodynamics calculations, and features its own time step subcycling algorithm in order to not slow down the hydrodynamics.  Changes in energy and composition are fed back into the SPH calculation at each (SPH) time step.  It accurately models the energy production during explosive burning to within 20\%.

\subsection{Detailed Nucleosynthetic Yields}

In order to obtain more accurate nucleosynthetic data, the thermodynamic trajectories of the particles were post-processed using the Burnf code \citep{YoungFryer07}.  Burnf is a flexible network (e.g., choice of isotopes, etc.) that employs the same architecture and microphysics as Tycho.  This work is focused on comparisons with species readily observable in supernova remnants.  With this aim, it was possible to economize on processor time by using a moderately sized network.  Calculations here used a 524-isotope network complete up to \iso{Tc}{99}, which provides accurate yields through the weak s-process. Reverse rates are calculated from detailed balance and allow a smooth transition to a nuclear statistical equilibrium (NSE) solver at temperatures $T > 10^{10}\ \mathrm{K}$. Neutrino cooling from plasma processes and the Urca process is calculated. For this work, Burnf chooses an appropriate time step based on the rate of change of abundances and performs a log-linear interpolation in the thermodynamic trajectory of each zone in the explosion calculation. The initial abundances are those of the 177 nuclei in the initial stellar model. Only particles that reached temperatures $T > \sci{2}{8}\ \mathrm{K}$ were post-processed. The individual particle yields from post-processing with Burnf were recombined with the particle temporal and spatial information for analysis and visualization. Note that \citet{Harris17} discuss some of the inherent issues and uncertainties involved in postprocessing abundances.

In addition to determining isotopic abundances for each individual SPH particle, we also recorded the peak temperature experienced by each particle over the length of the simulation and its density at the time of peak temperature.  The peak temperature and associated density are diagnostics of nucleosynthetic conditions in supernova explosions; they influence nuclear reaction rates, especially for the products of $\alpha$-rich freeze-outs, where rapidly changing conditions terminate nuclear reactions before NSE can be reached.  Production of \iso{Ti}{44} and \iso{Ni}{56}, and their resulting ratio can be very sensitive to peak temperatures, peak densities, and lepton fraction in the explosion \citep{Magkotsios10}, so we expect this ratio to be spatially correlated with those conditions.  High-energy emission lines from the decay of \iso{Ti}{44} can be detected in young remnants.
Under the assumption that most Fe-rich material originating in the interior of the remnant are dominated by \iso{Fe}{56} from \iso{Ni}{56} decay, the geometry of \iso{Ni}{56} production and abundance relative to \iso{Ti}{44} can be inferred \citep[e.g.,][]{grefenstette17}.



\section{\iso{Ti}{44} and \iso{Ni}{56} distribution}
\label{sec:tini}

The isotope \iso{Ti}{44} is produced in the innermost ejecta of a supernova and provides an ideal probe of the convective engine behind CCSNe.  Unlike \iso{Ni}{56}, the production of \iso{Ti}{44}, and even the path by which it is produced, depends on the exact conditions of the explosion.  \cite{Magkotsios10} identified a number of pathways for \iso{Ti}{44} that produce a wide variation in the ratio of \iso{Ti}{44} to \iso{Ni}{56}, including a ``QSE-leakage chasm'' where burning transitions from one QSE cluster to two, resulting in a mass fraction of \iso{Ti}{44} that is substantially lower than would be the case for slightly higher or lower peak temperatures.  Production of \iso{Ni}{56} is relatively insensitive to small changes in thermodynamic conditions. The strong sensitivity in the \iso{Ti}{44} production coupled to the much less sensitive \iso{Ni}{56} production make the ratio of these yields a strong probe of the details of the explosion mechanism.  In this section, we explore the dependency of the yields and, in particular, the ratio of \iso{Ti}{44} to \iso{Ni}{56}, on our specific explosion trajectories.

\begin{figure*}
\epsscale{1.15}
\plottwo{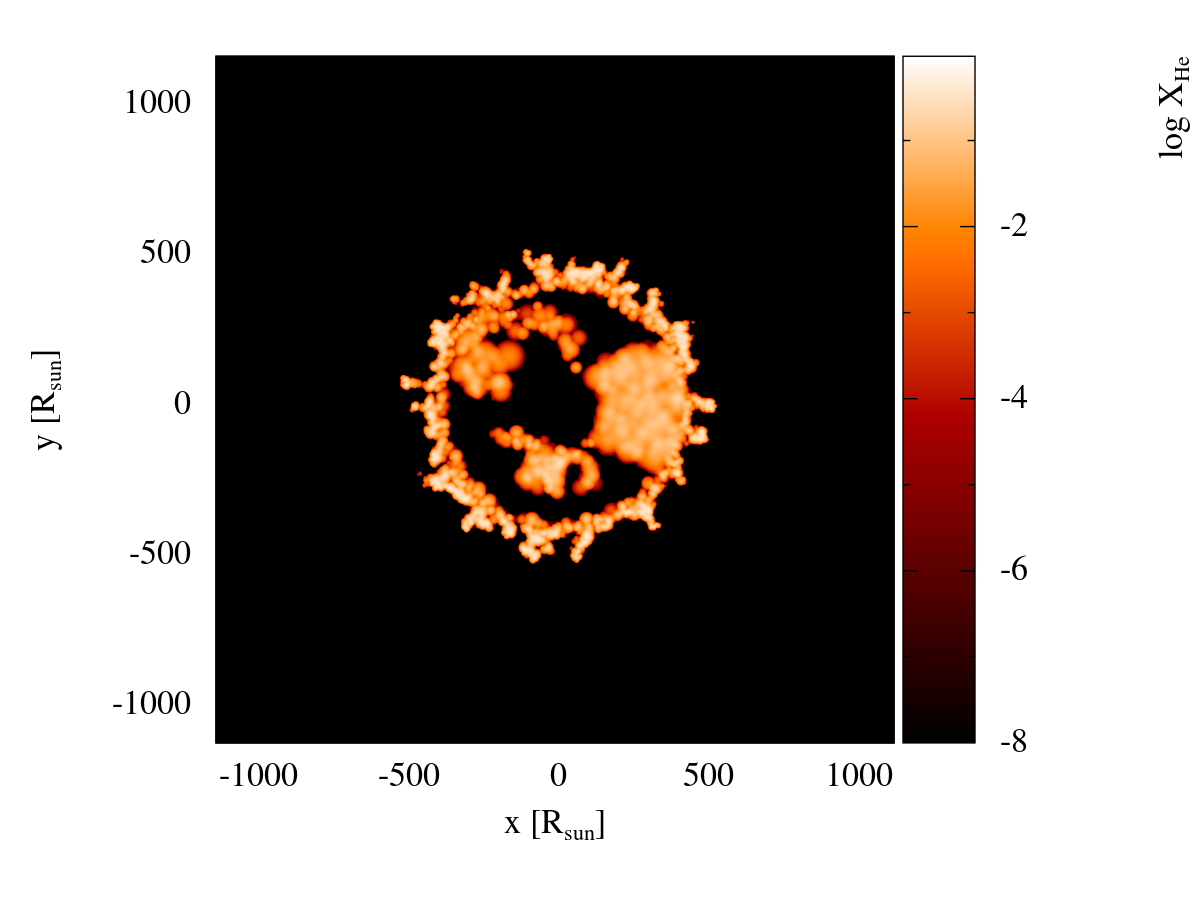}{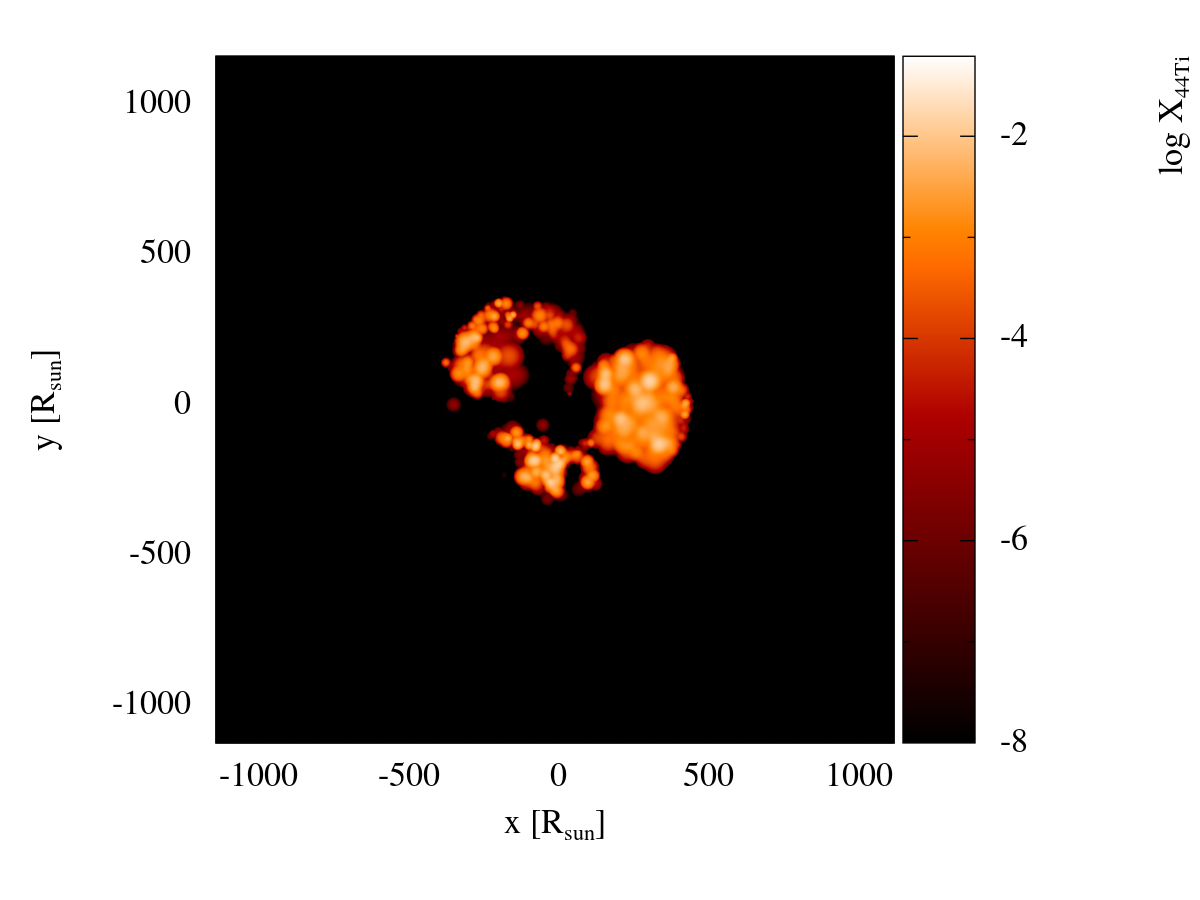}
\caption{Abundance of \iso{He}{4} (left) and \iso{Ti}{44} (right) in an $xy$ cross-section of the simulation.  Much of the \iso{Ti}{44} production corresponds to $\alpha$-rich regions in the ejecta.  Additional \iso{He}{4} is present in the Rayleigh-Taylor fingers formed by the reverse shock from the H-He interface at the base of the envelope traveling through products of partial He burning.}
\label{15M_dco_alpha_freezeout}
\end{figure*}

\begin{figure}
\epsscale{1.15}
\plotone{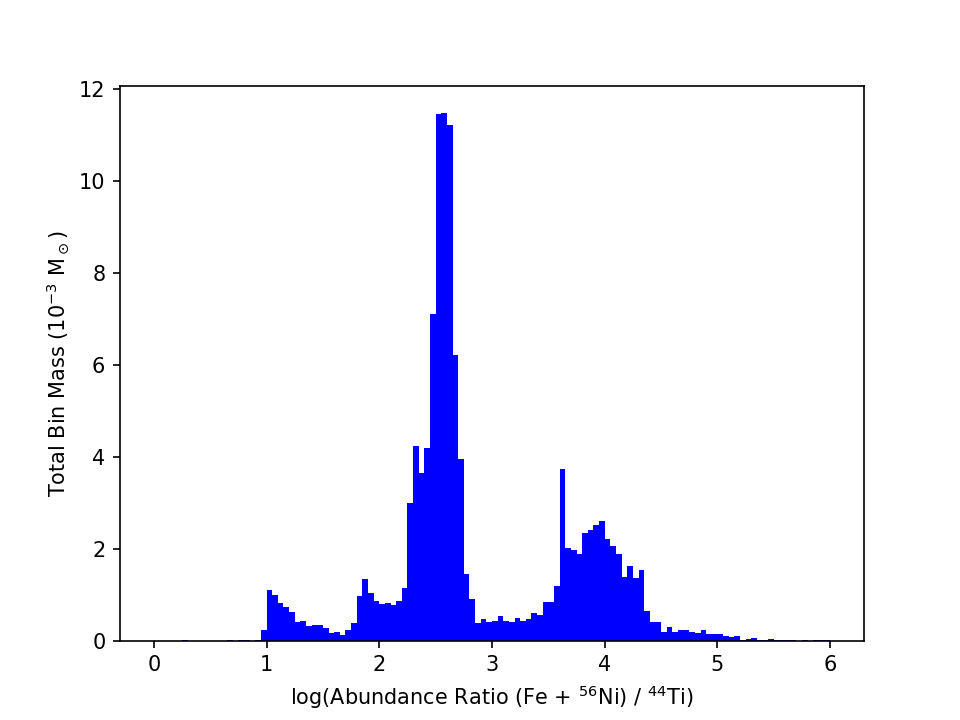}
\caption{Histogram of particle \feti{} ratios for the simulation.  Particles are weighted by their total masses, so each bin displays the sum of the masses of all particles it contains.}
\label{totalmass1dhist}
\end{figure}

\begin{figure}
\epsscale{1.15}
\plotone{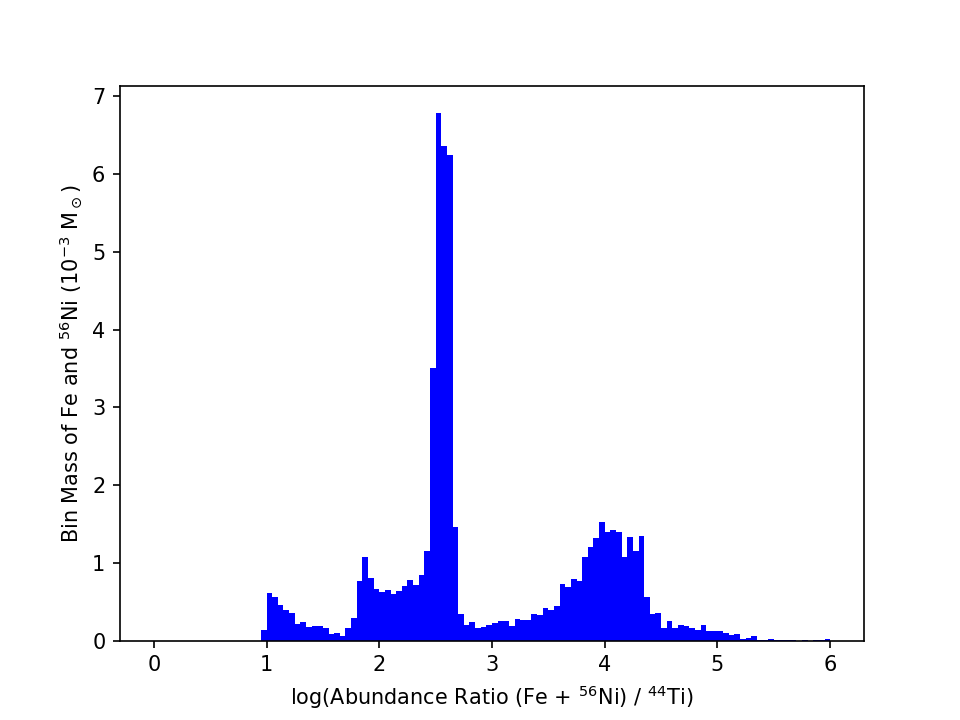}
\caption{Alternate histogram of particle \feti{} ratios for the simulation.  Particles are weighted by their total \emph{iron} masses, so each bin displays the sum of the masses of all Fe isotopes plus \iso{Ni}{56} from all particles in the bin.}
\label{ironmass1dhist}
\end{figure}

The nature of the convective engine is that it produces strong explosions along some directions (or lobes) with weaker explosions in between.  The $\alpha$-rich freeze-out region that \cite{Magkotsios10} argued would be the most efficient at producing \iso{Ti}{44} is tied to the strong explosive lobes and can be identified by regions where the \iso{He}{4} abundance fraction is also high.  Figure~\ref{15M_dco_alpha_freezeout}~(left) shows the \iso{He}{4} abundance in $xy$ cross-section of our three-dimensional model.  The corresponding \iso{Ti}{44} distribution, as shown in figure~\ref{15M_dco_alpha_freezeout}~(right), traces this $\alpha$-rich region.  Regions without a large fraction of $\alpha$ particles (\iso{He}{4} nuclei) did not produce much \iso{Ti}{44}.  The isotope \iso{Ni}{56} is produced in this same region, but the \iso{Ni}{56}/\iso{Ti}{44} ratio varies considerably.  Figures~\ref{totalmass1dhist} and \ref{ironmass1dhist} are histograms of all particle \feti{} values for the simulation for particles having \iso{Ti}{44} mass fraction $X(\iso{Ti}{44}) > 10^{-6}$. In figure~\ref{totalmass1dhist}, the particles are each weighted by their total mass. In figure~\ref{ironmass1dhist}, the particles are instead weighted by their total \emph{iron} mass (including all Fe isotopes and \iso{Ni}{56}, since \iso{Ni}{56} eventually decays to \iso{Fe}{56}).  Each histogram shows two distinct populations of particles: those having \feti{} ratios of $\sim 10^{2.5}$ and those having \feti{} ratios of $\sim 10^{3.9}$. There are also minor peaks visible with \feti{} ratios near $10^{1.0}$ and $10^{1.8}$.  

\begin{figure}
\epsscale{1.15}
\plotone{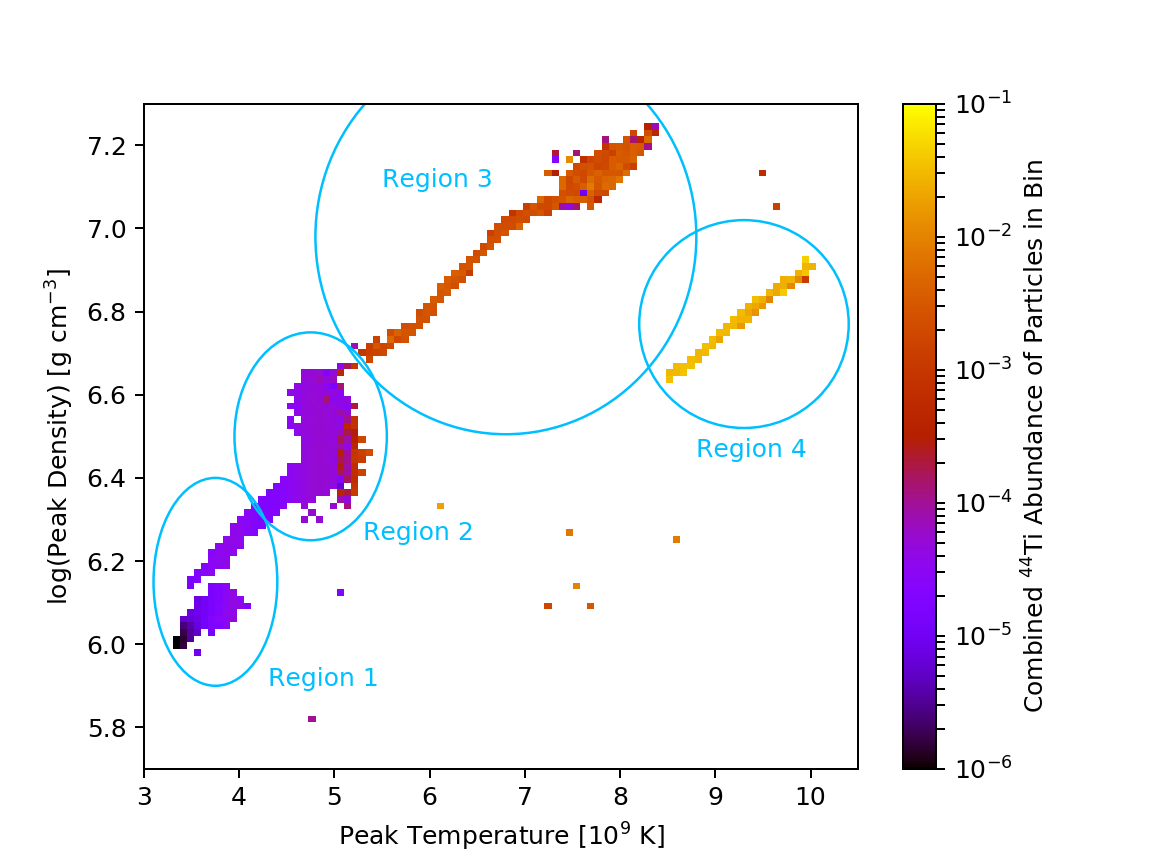}
\caption{Two-dimensional histogram of the simulation particles by peak temperature and associated density during the run. Each particle is weighted by its \iso{Ti}{44} mass divided by the total mass of all particles in the bin, so colors indicate the mass-averaged \iso{Ti}{44} abundance of all contained particles. This figure is produced in the style of \cite{Magkotsios10} for ease of comparison to their work.}
\label{magkotsiosplot}
\end{figure}

\begin{figure}
\epsscale{1.15}
\plotone{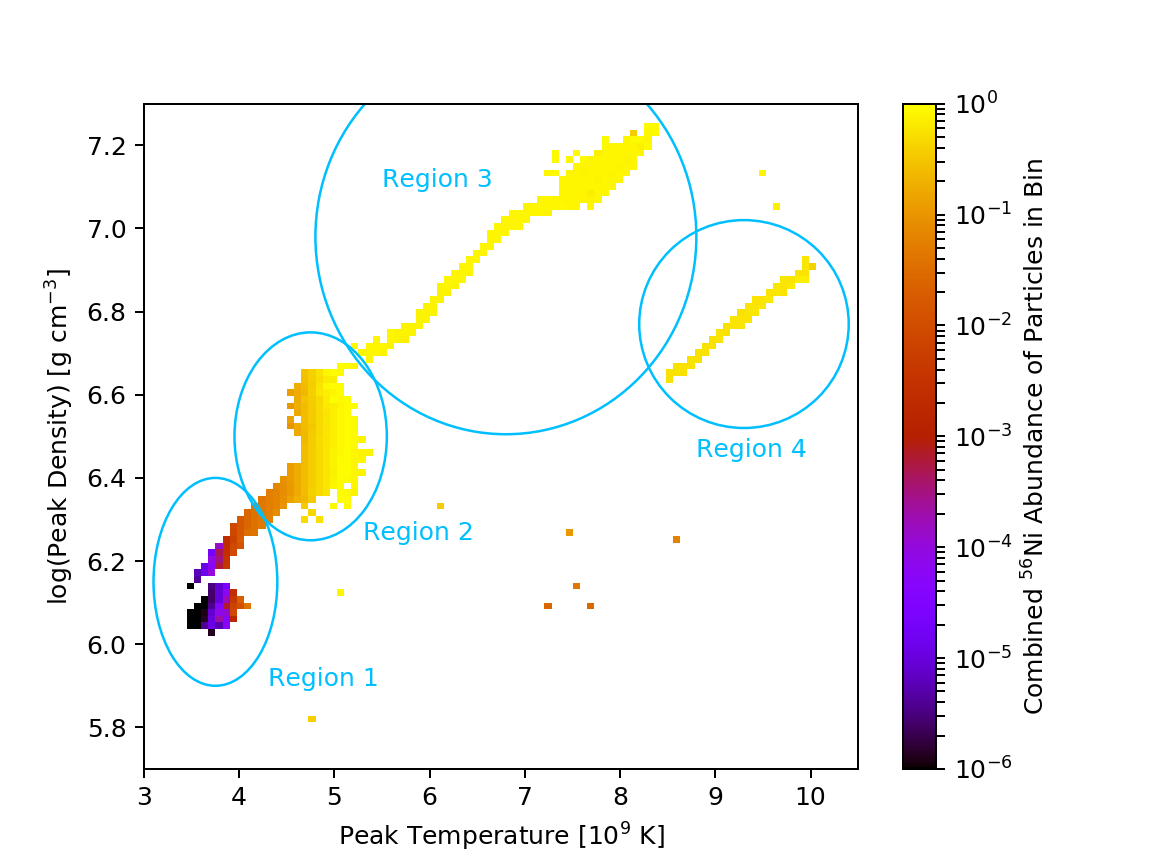}
\caption{Same as figure~\ref{magkotsiosplot}, but colors indicate the combined \iso{Ni}{56} abundance of each bin.}
\label{magkotsios_ni56}
\end{figure}

To understand the variations in the \feti{} ratio better, we must explore how the temperature and density evolution of different ejecta can influence the final \iso{Ti}{44} yield.  \cite{Magkotsios10} found that the peak temperature and density (at time of peak temperature) of material ejected from a supernova dictate the final \iso{Ti}{44} and \iso{Ni}{56} yields.  While \cite{Magkotsios10} studied a wide range of peak temperature/density pairs, we focus on the pairs encountered for the specific trajectories in our simulation.  Figure~\ref{magkotsiosplot} shows the \iso{Ti}{44} abundance of our SPH particles as a function of the peak explosion conditions.  This figure was produced by binning all the simulation particles in two dimensions by their peak temperature and associated density during the run.  The color displayed in each bin indicates the total \iso{Ti}{44} mass fraction $X_\mathrm{bin} (\iso{Ti}{44})$ of particles in the bin, which is given by
\begin{equation}
X_\mathrm{bin} (\iso{Ti}{44}) = \frac{\sum_{i \in \mathrm{bin}} m_i X_i (\iso{Ti}{44})}{\sum_{i \in \mathrm{bin}} m_i} ,
\end{equation}
where $X_i$ denotes the mass fraction of a particular isotope or element in particle $i$, $m_i$ is the total mass of particle $i$, and summation over $i \in \mathrm{bin}$ means summing for all particles $i$ in the bin.  Figure~\ref{magkotsiosplot} is scaled and colored in the style of \cite{Magkotsios10} for ease of comparison with their results.  The corresponding plot for \iso{Ni}{56} production is shown in figure~\ref{magkotsios_ni56}.  Comparing these plots gives a first look at the conditions that produce the lowest and highest \feti ratios.  The hotter trajectories are the most efficient at producing \iso{Ni}{56} and \iso{Ti}{44}.


\begin{figure}
\epsscale{1.15}
\plotone{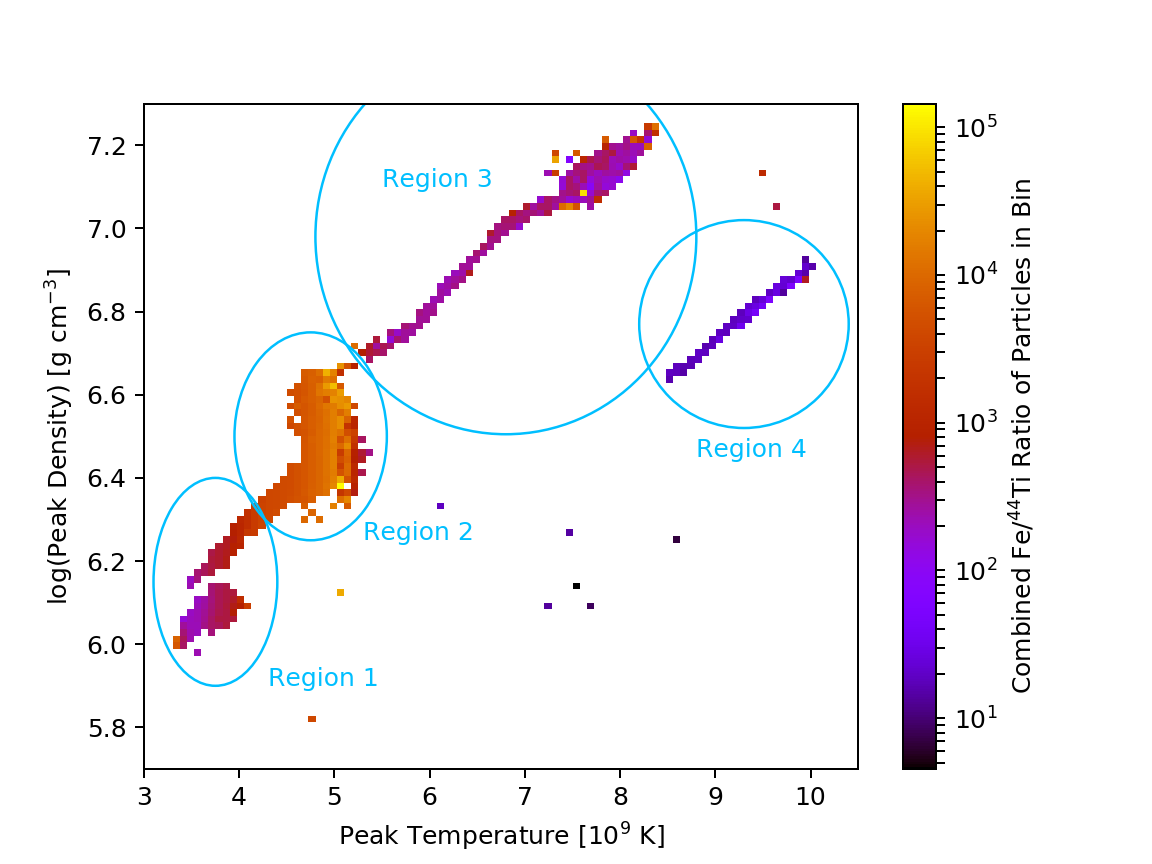}
\caption{Two-dimensional histogram similar to figures~\ref{magkotsiosplot} and \ref{magkotsios_ni56}, but with each bin color indicating the aggregate mass-weighted \feti{} ratio (including \iso{Ni}{56}) of all particles in the bin.}
\label{magkotsiosratio}
\end{figure}

Figure~\ref{magkotsiosratio} is similar to figures~\ref{magkotsiosplot} and \ref{magkotsios_ni56}, but instead of displaying one abundance in each bin, the colors now indicate the aggregate mass-weighted \feti{} ratio of all particles in the bin, which is given by
\begin{equation}
\frac{X_\mathrm{bin} (\mathrm{Fe}) + X_\mathrm{bin} (\iso{Ni}{56})}{X_\mathrm{bin} (\iso{Ti}{44})} = \frac{\sum_{i \in \mathrm{bin}} m_i \left[ X_i (\mathrm{Fe}) + X_i (\iso{Ni}{56}) \right]}{\sum_{i \in \mathrm{bin}} m_i X_i (\iso{Ti}{44})} .
\end{equation}
This series of figures shows the temperature and density conditions under which \iso{Ti}{44} and \iso{Ni}{56} production result in the bimodal ratios that appear to be characteristic of both Cas~A and our simulation. Material produced at high temperatures in the $\alpha$-rich freezeout region has a very low \feti{} ratio, of order a few hundred. This transitions rapidly to \feti{} of several thousand or even several times $10^4$ at lower temperatures. At even lower temperatures, there is another rapid transition to a small population of particles with \feti{} of several hundred. In this regime, \iso{Ti}{44} and \iso{Ni}{56} are produced in similar amounts at mass fractions of order $10^{-5}$ by explosive silicon burning. The \feti{} ratio is still several hundred due to the presence of iron isotopes remaining from the progenitor star's initial composition. 

\begin{figure*}
\epsscale{1.17}
\plottwo{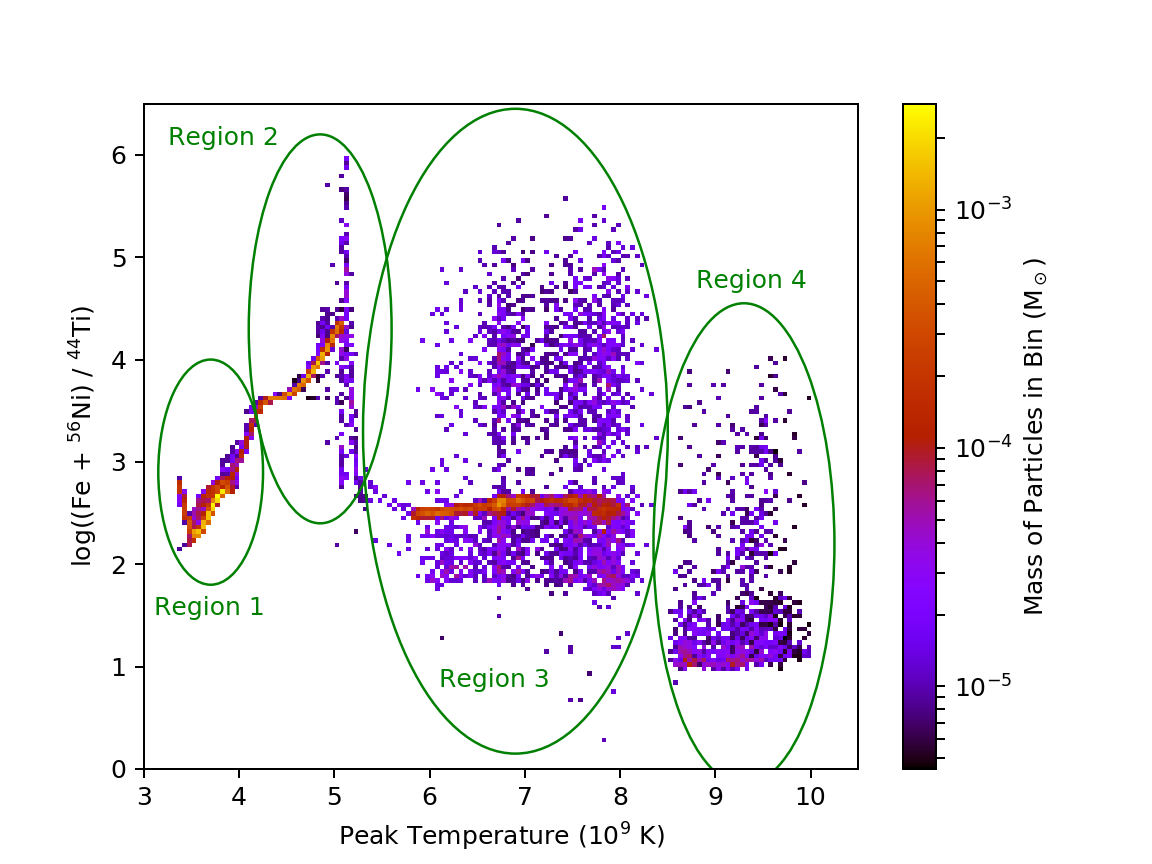}{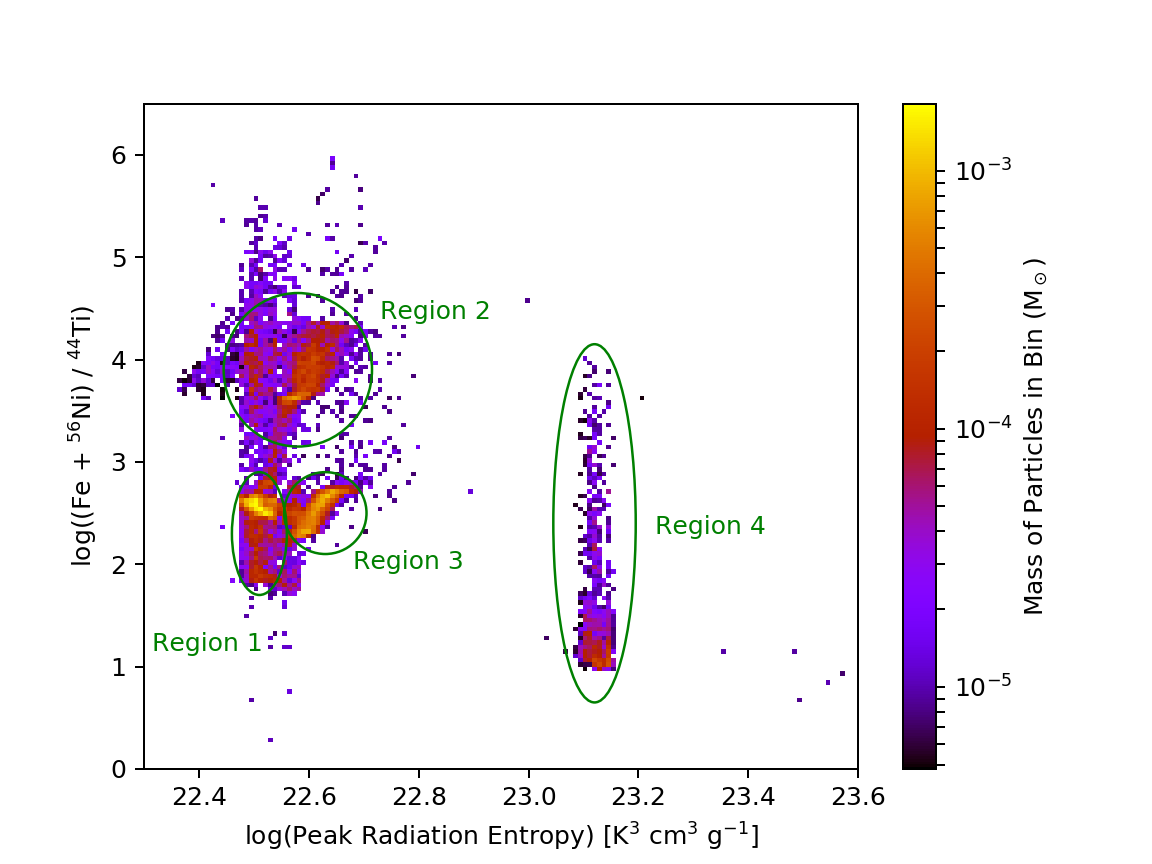}
\caption{Two-dimensional histograms of \feti{} ratio as a function of peak temperature (left) and peak radiation entropy (right).  The colors indicate the sum of the particle masses in each bin.  Regions 1, 2, 3, and 4 (appearing from left to right) are well-separated in peak temperature.}
\label{feti-T-SR}
\end{figure*}

Figure~\ref{feti-T-SR} shows two-dimensional histograms of \feti{} ratio plotted against peak temperature and peak radiation entropy, where the total mass in each bin is indicated by the color scales.  Peak radiation entropy is defined as the radiation entropy at the time of peak temperature, which is related to the peak temperature and density via
\begin{equation}
\srad (T_\mathrm{peak}) = \frac{T_\mathrm{peak}^3}{\rho (T_\mathrm{peak})} ,
\end{equation}
where $T_\mathrm{peak}$ is the peak temperature and $\rho (T_\mathrm{peak})$ is the associated density.  It can be seen that region 4, which has very low typical values of \feti{}, represents a small amount of overall mass relative to the other regions.  The regions are well separated in peak temperature space, but the separation is much less clear in radiation entropy, since there is considerable degeneracy in $T^3/\rho$ pairs that result in the same value of \srad.

Employing these plots, we can now better understand the yields from this explosion.  To this end, we have separated the ejecta into four regions in peak temperature/density space that are numbered 1 through 4.  Table~\ref{stats_table} provides precise definitions for the boundaries delimiting each region, as well as some thermodynamic quantities for the mean thermodynamic trajectories of particles in each.  \cite{Magkotsios10} identified multiple burning regions, and the yields from our first two regions can be understood by comparing to these burning regions.

\begin{deluxetable*}{ccccccc}
\tablewidth{0pt}
\tablecaption{Thermodynamic properties for the mean trajectories of each region.  The second and third columns delimit the extent used in defining each region of particles. \label{stats_table}}
\tablehead{\colhead{Region} & \colhead{$T_\mathrm{peak}$} & \colhead{$\log \rho(T_\mathrm{peak})$} & \colhead{$T_\mathrm{peak}$} & \colhead{$\log \rho(T_\mathrm{peak})$} & \colhead{$\log S_\mathrm{rad}(T_\mathrm{peak})$} & \colhead{$\log(\feti)$} \\
\colhead{Number} & \colhead{Range} & \colhead{Range} & \colhead{Mean} & \colhead{Mean} & \colhead{Mean} & \colhead{Mean} \\
\colhead{---} & \colhead{($10^9$ K)} & \colhead{$[\mathrm{g\ cm^{-3}}]$} & \colhead{($10^9$ K)} & \colhead{$[\mathrm{g\ cm^{-3}}]$} & \colhead{$[\mathrm{K^3\ cm^3\ g^{-1}}]$} & \colhead{---}} 
\startdata
1 & (3.3, 4.2) & (5.9, 6.4) & 3.65 & 6.08 & 22.61 & 2.58 \\
2 & (4.2, 5.3) & (6.1, 6.8) & 4.77 & 6.45 & 22.59 & 3.92 \\
3 & (5.3, 8.4) & (6.6, 7.3) & 7.04 & 7.03 & 22.51 & 2.82 \\
4 & (8.4, 10.1) & (6.5, 7.0) & 9.20 & 6.77 & 23.12 & 1.57 \\
\enddata
\end{deluxetable*}

Region 1 includes the outermost ejecta, the coldest material that still produces \iso{Ti}{44}.  The conditions in this ejecta correspond to the ``Si-rich'' zone (incomplete burning) identified by \cite{Magkotsios10} to the left of the Ti-depleted chasm. Nuclear burning in this region produces and destroys \iso{Ti}{44} through a variety of pathways and is sensitive to many nuclear reactions.

Region 2 is slightly hotter and corresponds to material just to the right of the QSE leakage chasm.  The burning is more complete and the \iso{Ni}{56} production rises dramatically, but it is still sensitive to some rates and the exact conditions of the ejecta.   \cite{Magkotsios10} found that, for the same temperature, the \iso{Ti}{44} production is lower at higher densities (as it approaches the QSE leakage chasm) and peaks at slightly lower densities than those found in our model.  In this way, the \iso{Ti}{44} abundance is sensitive to the progenitor: the higher the progenitor's mass, the higher the density in the trajectories producing a lower \iso{Ti}{44} yield in this region.

Region 3 corresponds to $\alpha$-rich nucleosynthesis produced in the energetic outflows of our asymmetric ejecta.  This material is the least sensitive to uncertainties in nuclear rates.

Region 4 is characterized by very hot (above $\sim \sci{9}{9}\ \mathrm{K}$) but relatively low-density material.  These conditions are very efficient at producing \iso{Ti}{44} and this ejecta is responsible for the small amount of material appearing in figure~\ref{ironmass1dhist} near $\log(\feti{}) \sim 1.0$ and 1.8.  In general, our yields differ only slightly from those of \cite{Magkotsios10}, but we have much more efficient \iso{Ti}{44} production in our region 4 than a peak temperature/density solution based on the models of \cite{Magkotsios10}.  This material is the highest-entropy ejecta caused by the strongest explosion lobes.  This is in the $\alpha$p-rich region identified by \cite{Magkotsios10}.  In \cite{Magkotsios10}, this region produced \iso{Ti}{44} with the same efficiency as the $\alpha$-rich region.  In our simulations, this ejecta very efficiently produces \iso{Ti}{44}, yielding the lowest \feti{} ratios.

\begin{figure}
\epsscale{1.15}
\plotone{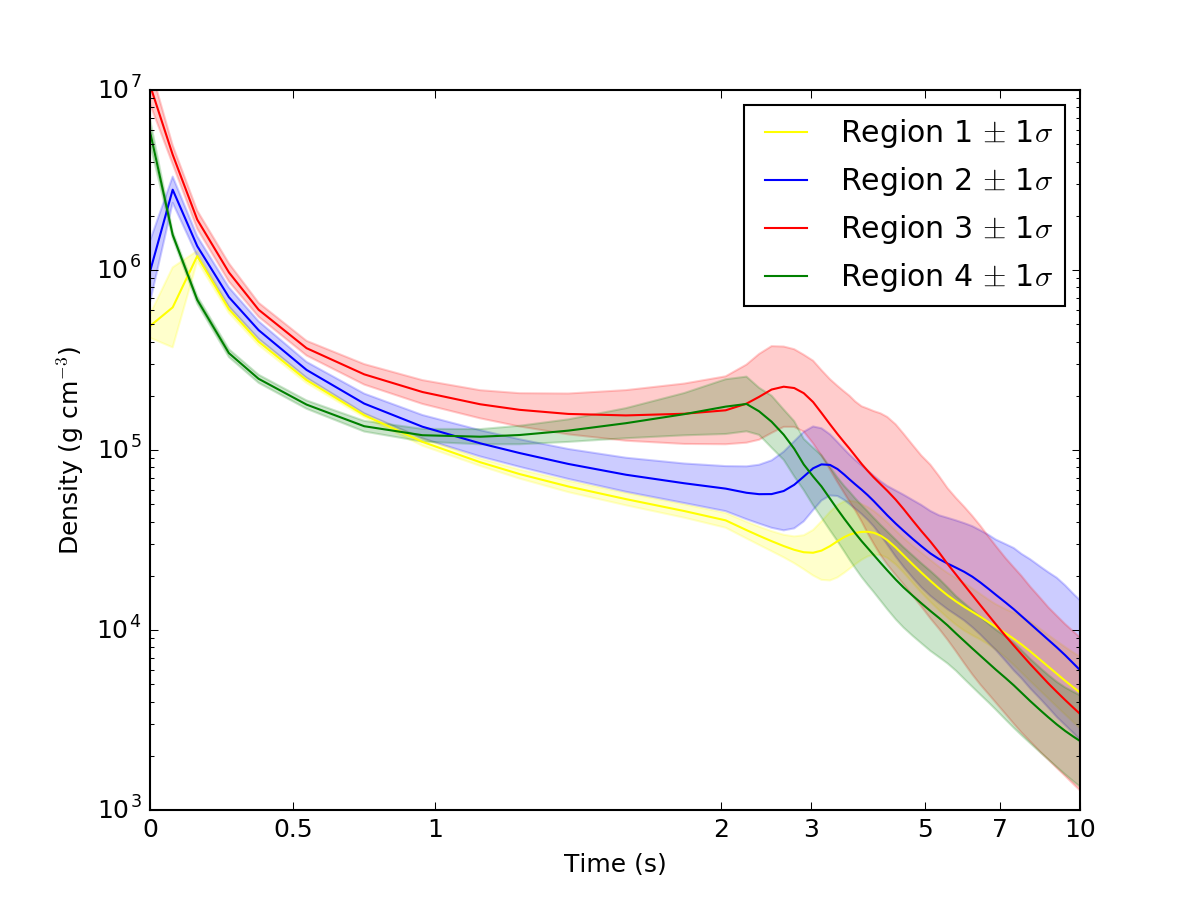}
\caption{Mean density of thermodynamic trajectories for the \iso{Ti}{44}-rich regions (region 1 in yellow, region 2 in blue, region 3 in red, and region 4 in green). Shaded areas represent $1\sigma$ spreads for particle densities. Means and standard deviations were calculated in log space. The $t$ axis is a ``symmetric log scale,'' which is linear near zero to more effectively show values at early times.}
\label{t-rho-sigma}
\end{figure}

\begin{figure}
\epsscale{1.15}
\plotone{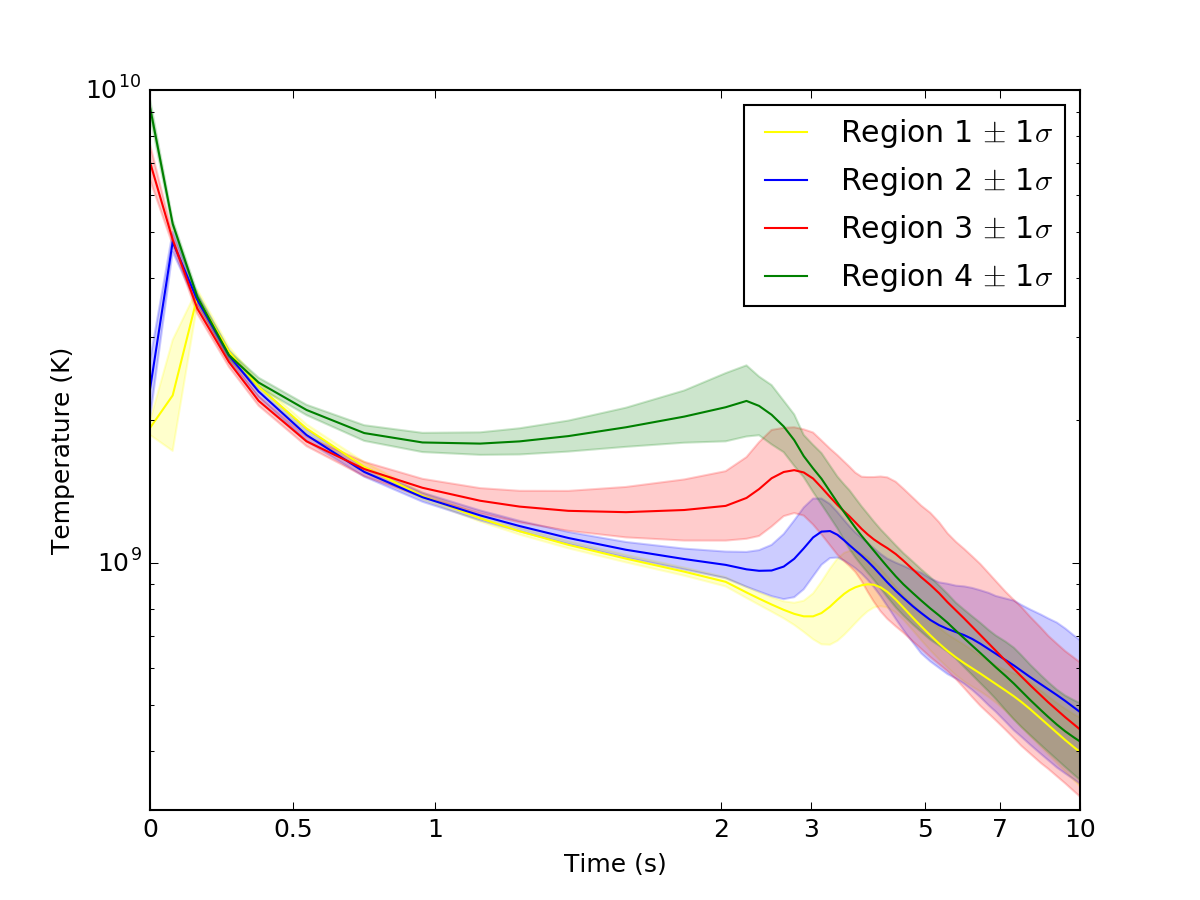}
\caption{Mean temperature of thermodynamic trajectories for the \iso{Ti}{44}-rich regions (region 1 in yellow, region 2 in blue, region 3 in red, and region 4 in green). Shaded areas represent $1\sigma$ spreads for particle temperatures. Means and standard deviations were calculated in log space. The $t$ axis is a ``symmetric log scale,'' which is linear near zero to more effectively show values at early times.}
\label{t-T-sigma}
\end{figure}

\begin{figure}
\epsscale{1.15}
\plotone{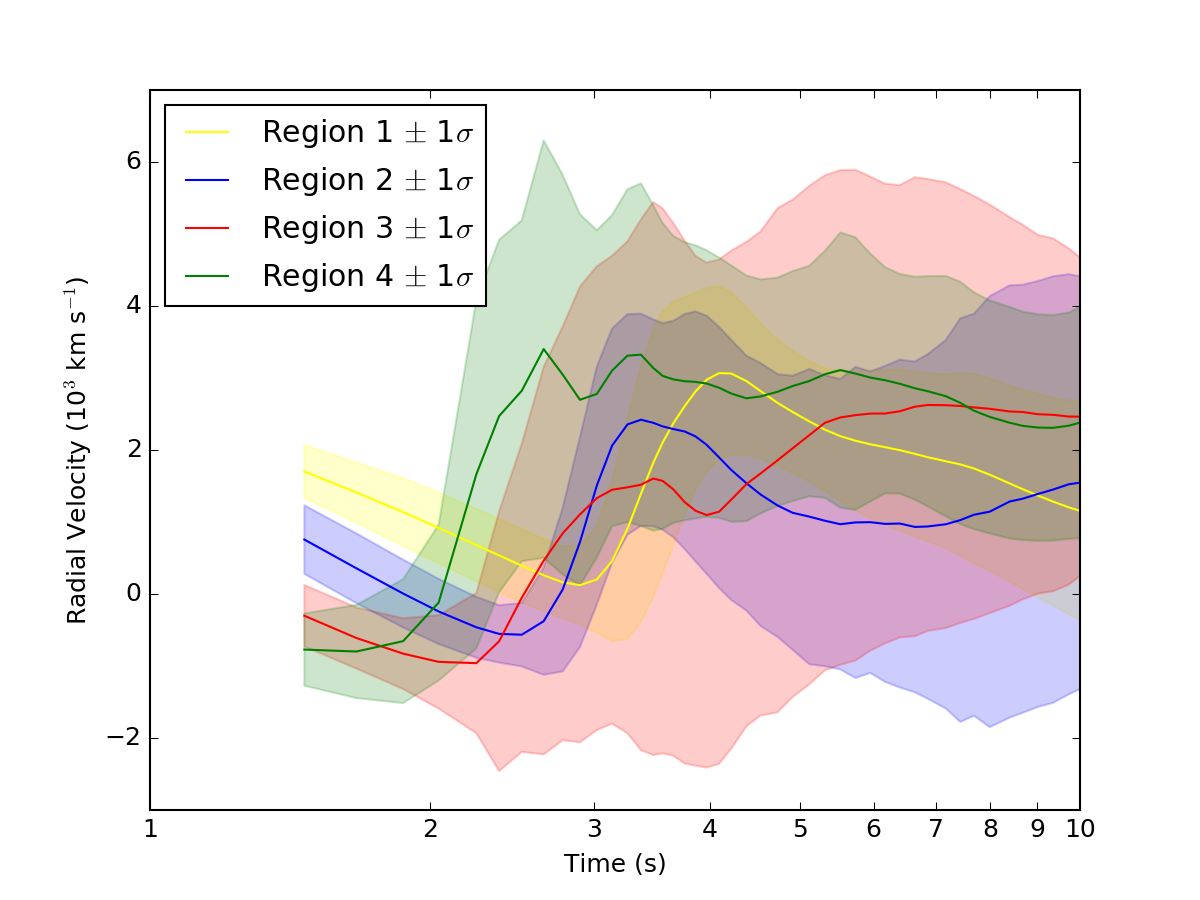}
\caption{Mean radial velocity of thermodynamic trajectories for the \iso{Ti}{44}-rich regions (region 1 in yellow, region 2 in blue, region 3 in red, and region 4 in green). Shaded areas represent $1\sigma$ spreads for particle radial velocities. Unlike figures~\ref{t-rho-sigma} and \ref{t-T-sigma}, means and standard deviations in this figure are calculated normally (i.e., in linear space). Also, the $t$ axis here uses a conventional log scale. Note: radial velocity data was unavailable for simulation times earlier than 1.25~s.}
\label{t-rvel-sigma}
\end{figure}

In the rest of this section, we will study these differences, determining that they arise from deviations of the evolution of the matter away from a simple power law or exponential trajectory.  Although these simple trajectories are useful for parameter studies like those of \cite{Magkotsios10}, the sensitivity of \iso{Ti}{44} production to the thermodynamic trajectory can make it difficult to produce exact yields with analytic trajectories. Figures~\ref{t-rho-sigma}--\ref{t-rvel-sigma} show the evolution of mean density, temperature, and radial velocity with time for the regions of interest. The lightly shaded regions show the $1\sigma$ spread of those quantities around the mean.

After times of 3--5~s, the shock stops accelerating and, in some cases, begins to decelerate. Convective motions result in some material with negative radial velocities at certain points in the evolution. This causes a pileup of matter, and the densities and temperatures of the ejecta's thermal trajectories tend to increase.  It is this pileup that alters the yields, causing differences when compared to expectations based on studies using power law trajectories.  Especially for our hottest ejecta, the pileup pushes the temperature into a regime where nuclear burning is extremely active, and it is unsurprising that the \iso{Ti}{44} production varies dramatically for this matter with respective to the \cite{Magkotsios10} results for the same peak temperatures and densities.  

\begin{figure*}
\epsscale{1.14}
\plottwo{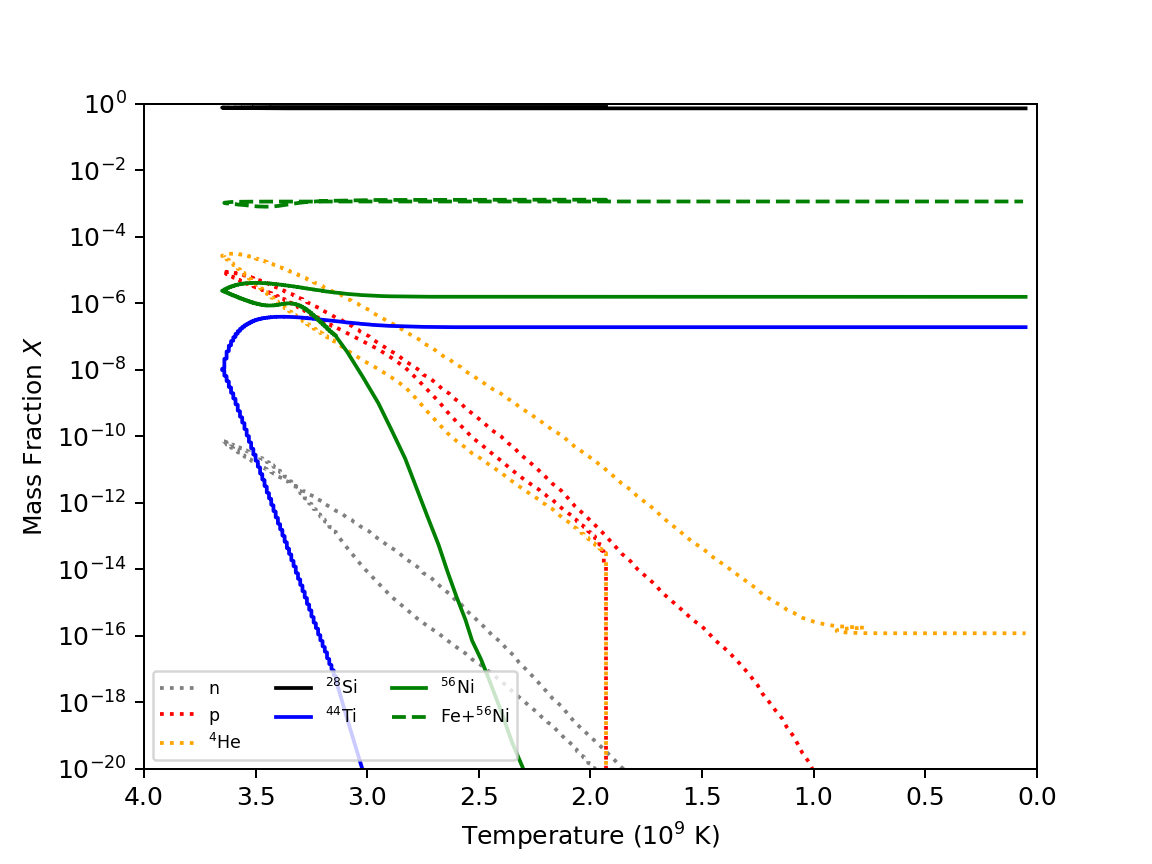}{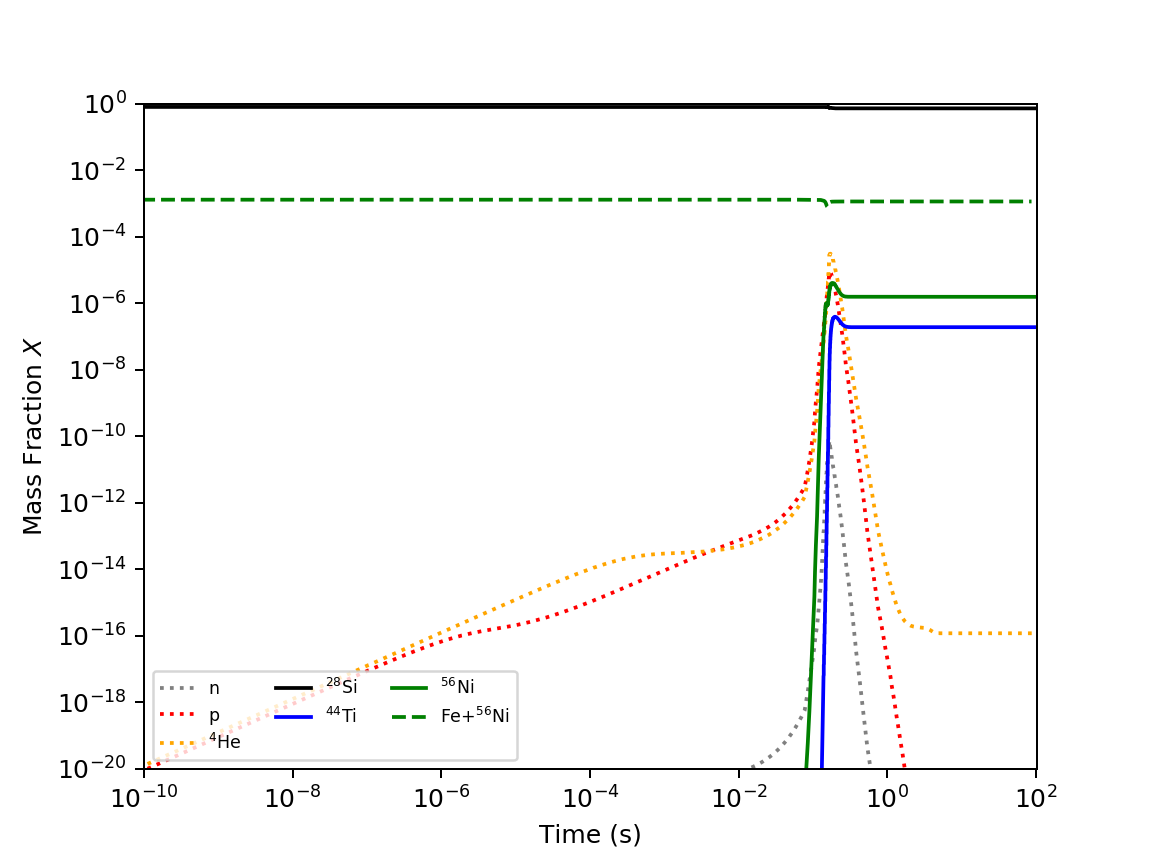}
\caption{Mass fraction evolution plotted against temperature (left) and time (right) for a series of significant isotopes. Temperature and density trajectories used were the mean of the region 1 particles. (See figures~\ref{t-rho-sigma} and \ref{t-T-sigma}.)}
\label{traj_reg1}
\end{figure*}

\begin{figure*}
\epsscale{1.14}
\plottwo{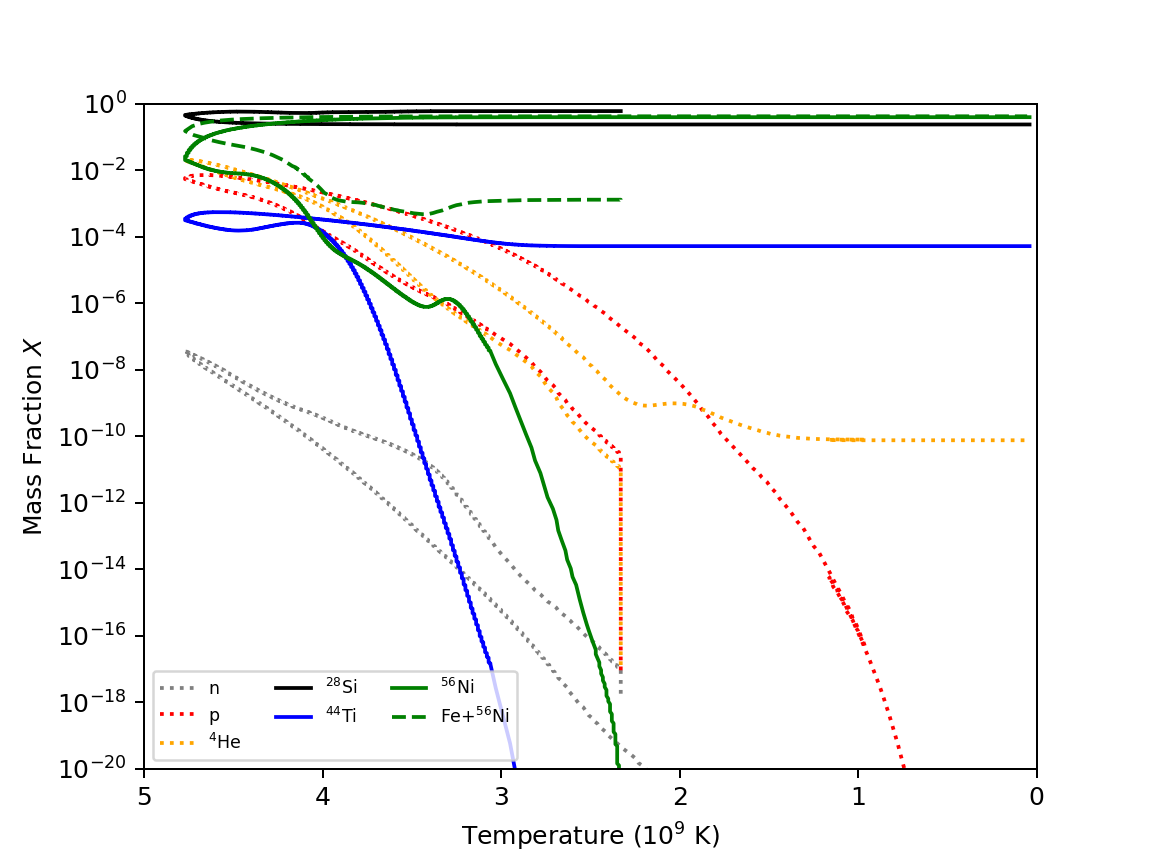}{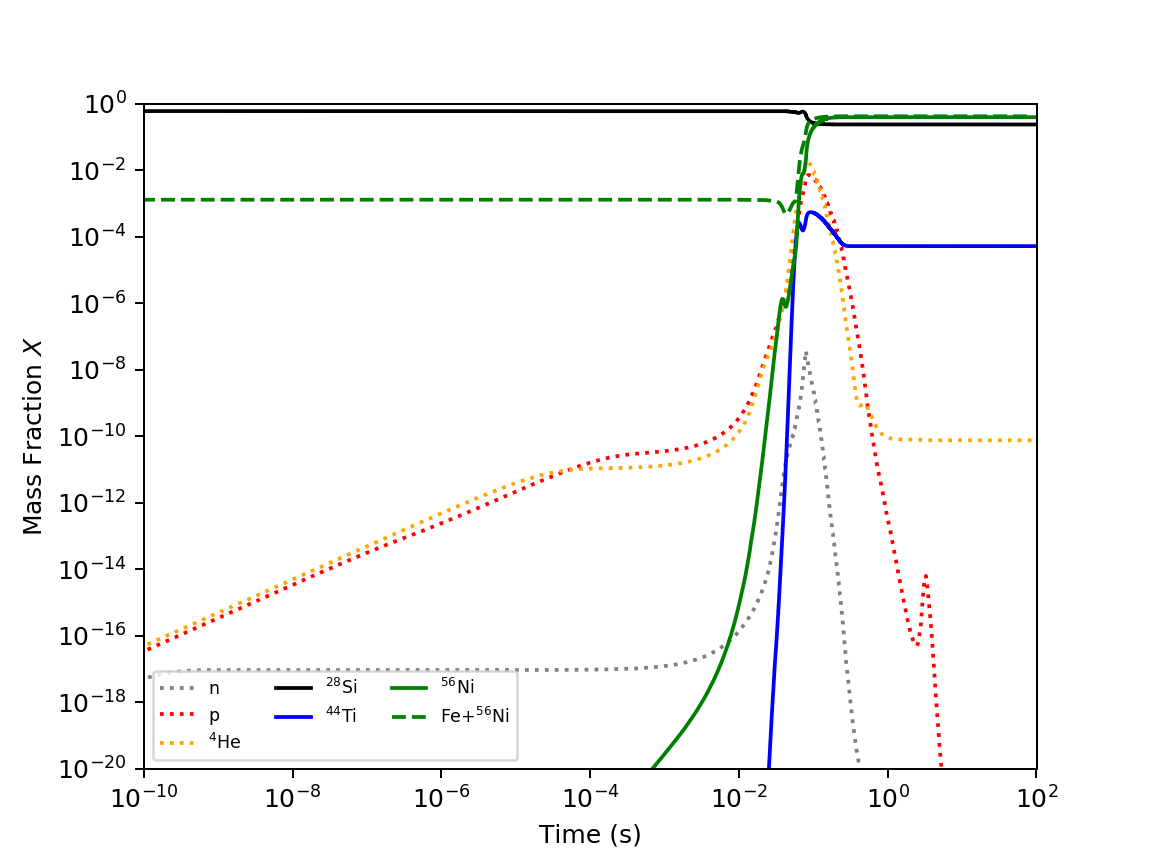}
\caption{Same as Figure~\ref{traj_reg1}, but for the region 2 particles' mean trajectories.}
\label{traj_reg2}
\end{figure*}

\begin{figure*}
\epsscale{1.14}
\plottwo{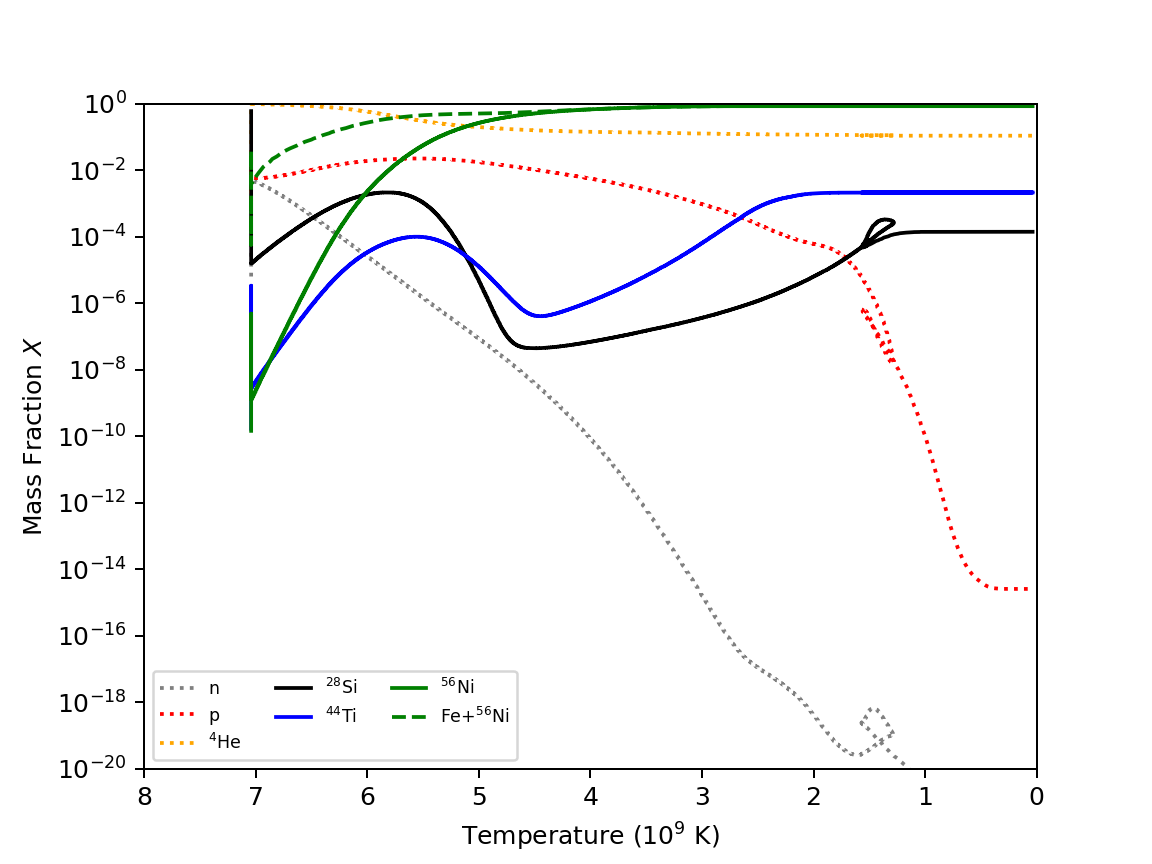}{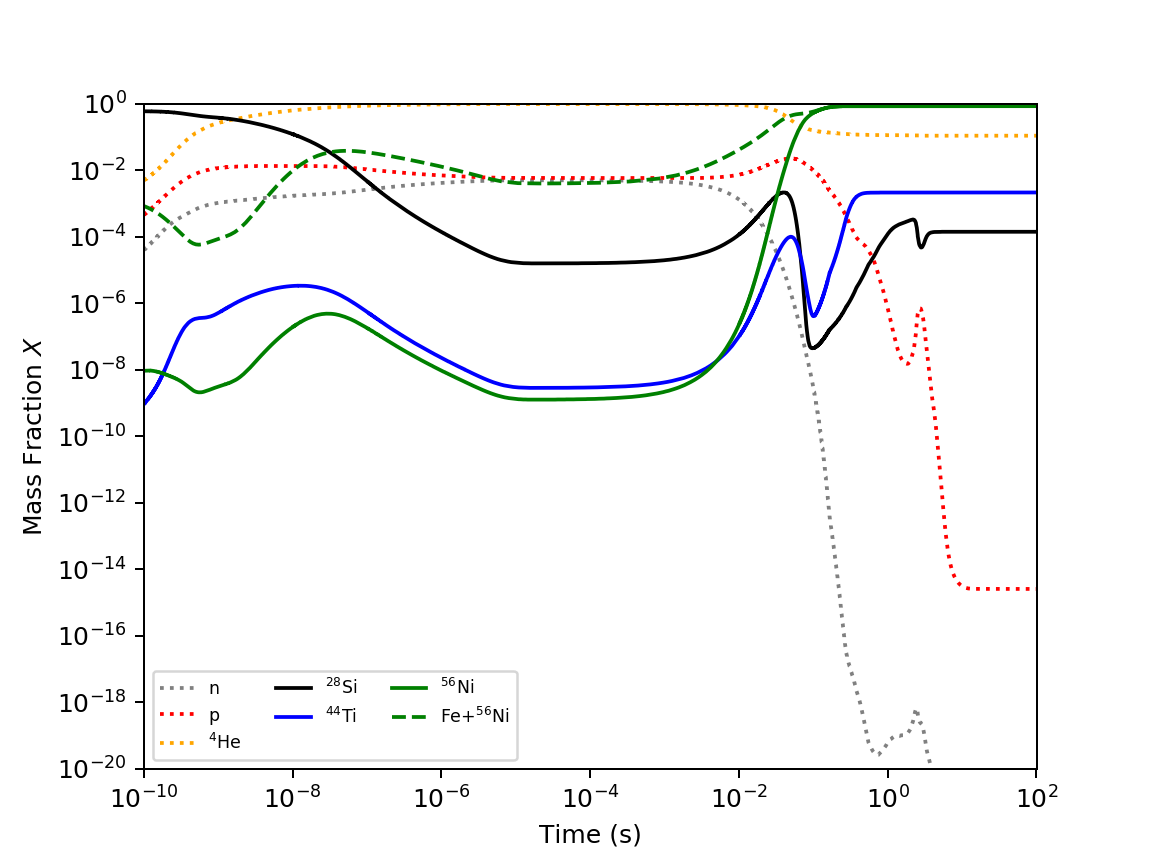}
\caption{Same as Figure~\ref{traj_reg1}, but for the region 3 particles' mean trajectories.}
\label{traj_reg3}
\end{figure*}

\begin{figure*}
\epsscale{1.14}
\plottwo{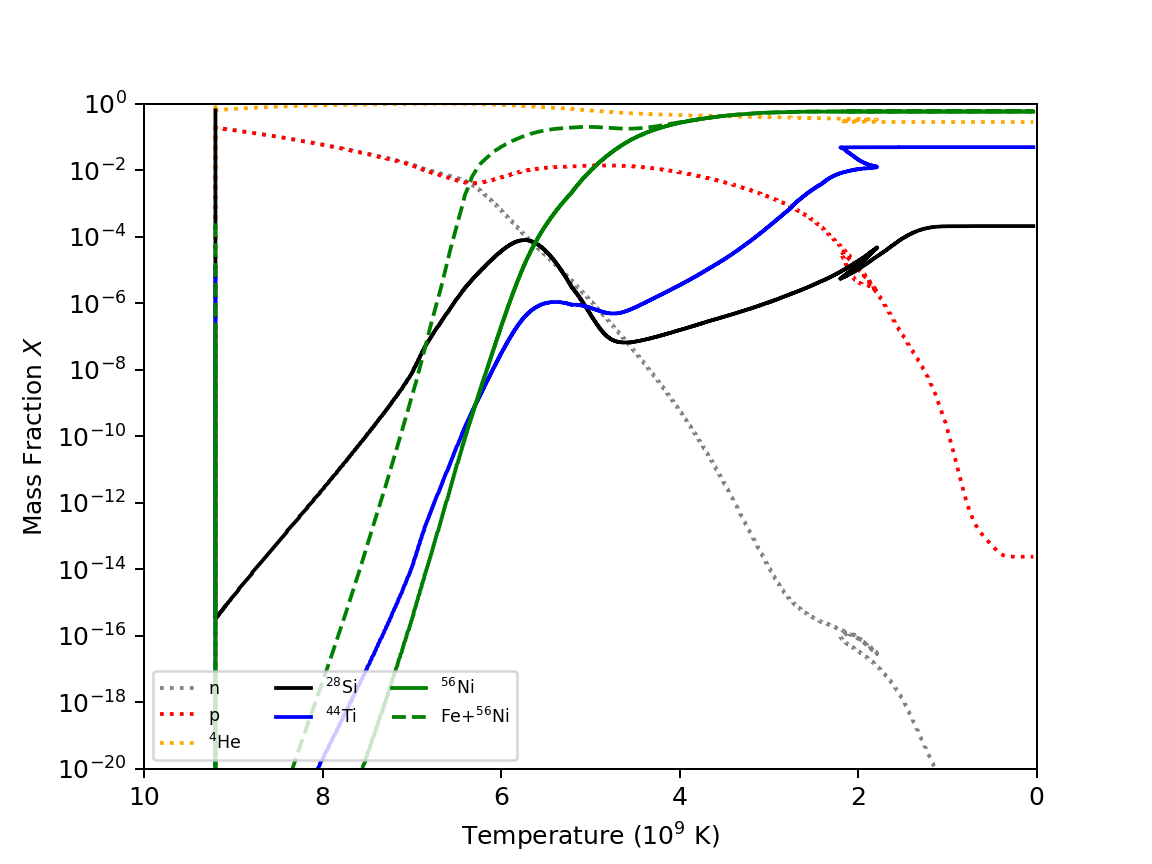}{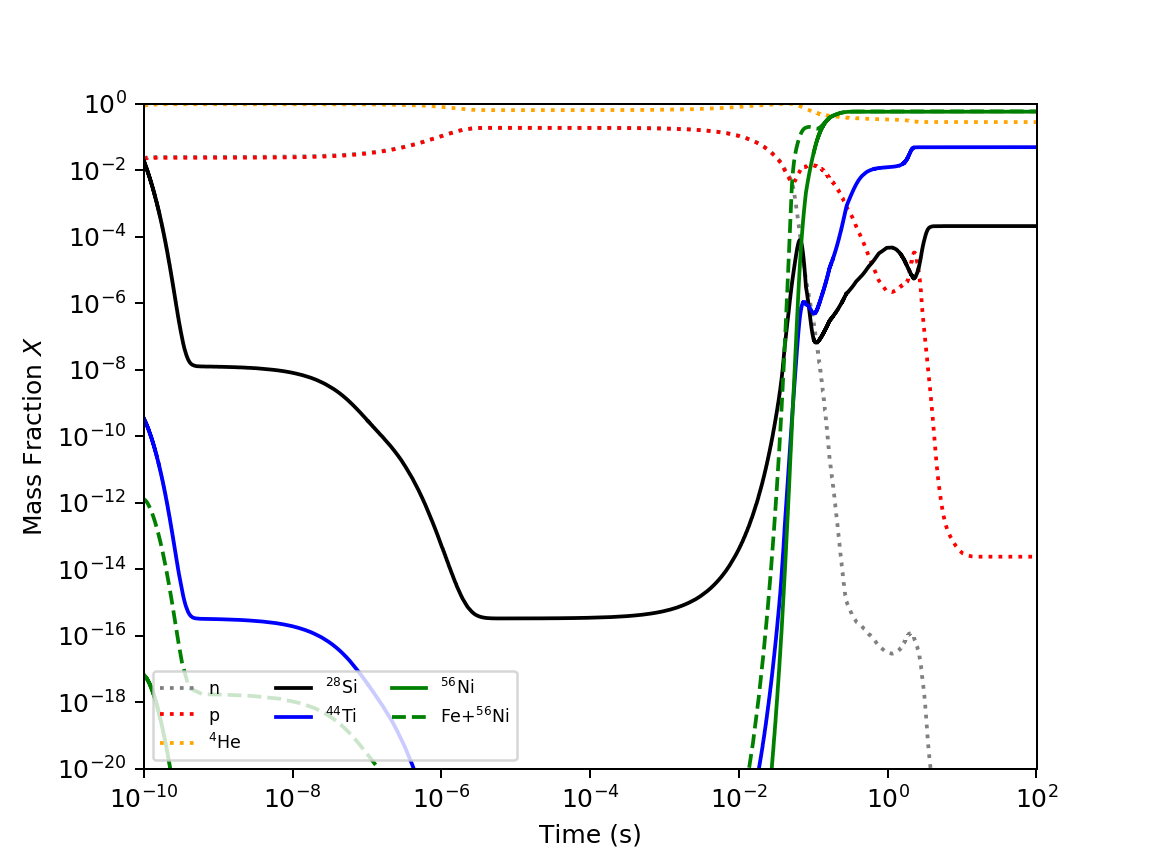}
\caption{Same as Figure~\ref{traj_reg1}, but for the region 4 particles' mean trajectories.}
\label{traj_reg4}
\end{figure*}

\begin{figure*}
\epsscale{1.14}
\plottwo{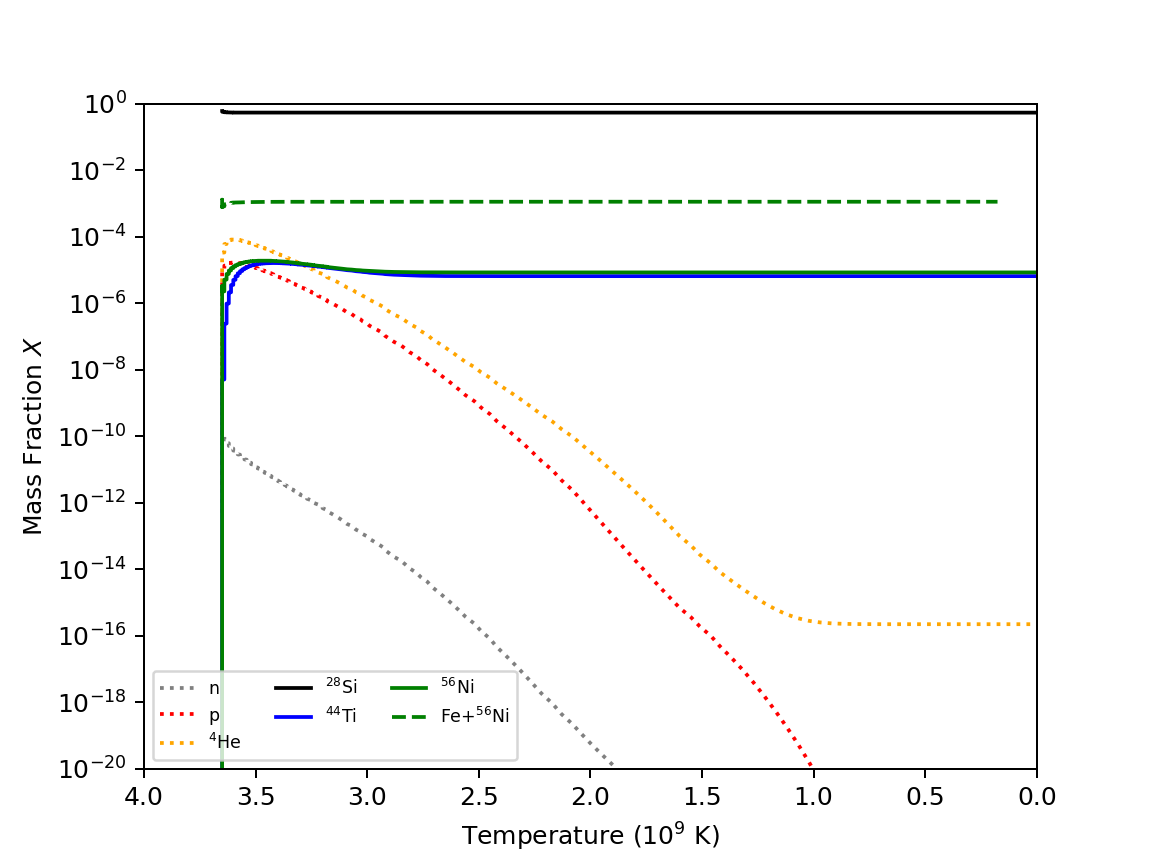}{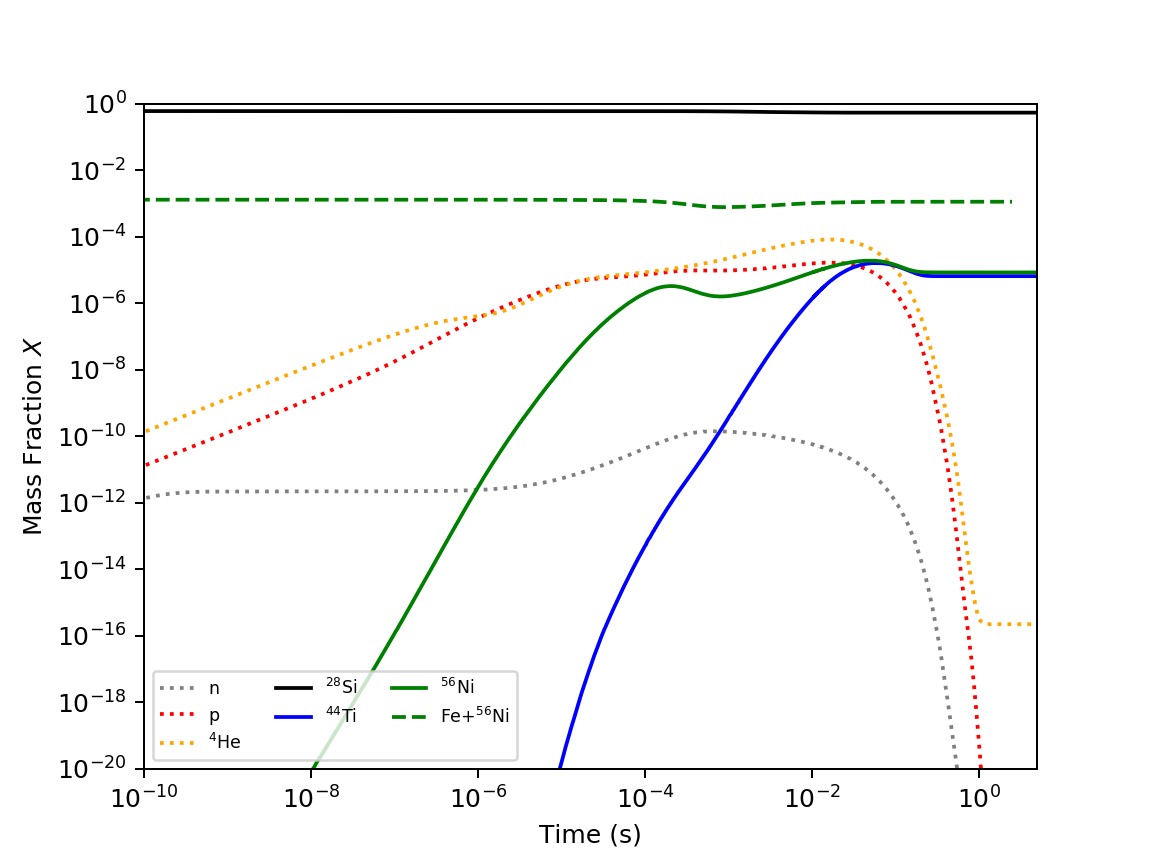}
\caption{Mass fraction evolution plotted against temperature (left) and time (right) for a series of significant isotopes using best-fit power law trajectories for temperature and density of the region 1 particles.}
\label{plaw_reg1}
\end{figure*}

\begin{figure*}
\epsscale{1.14}
\plottwo{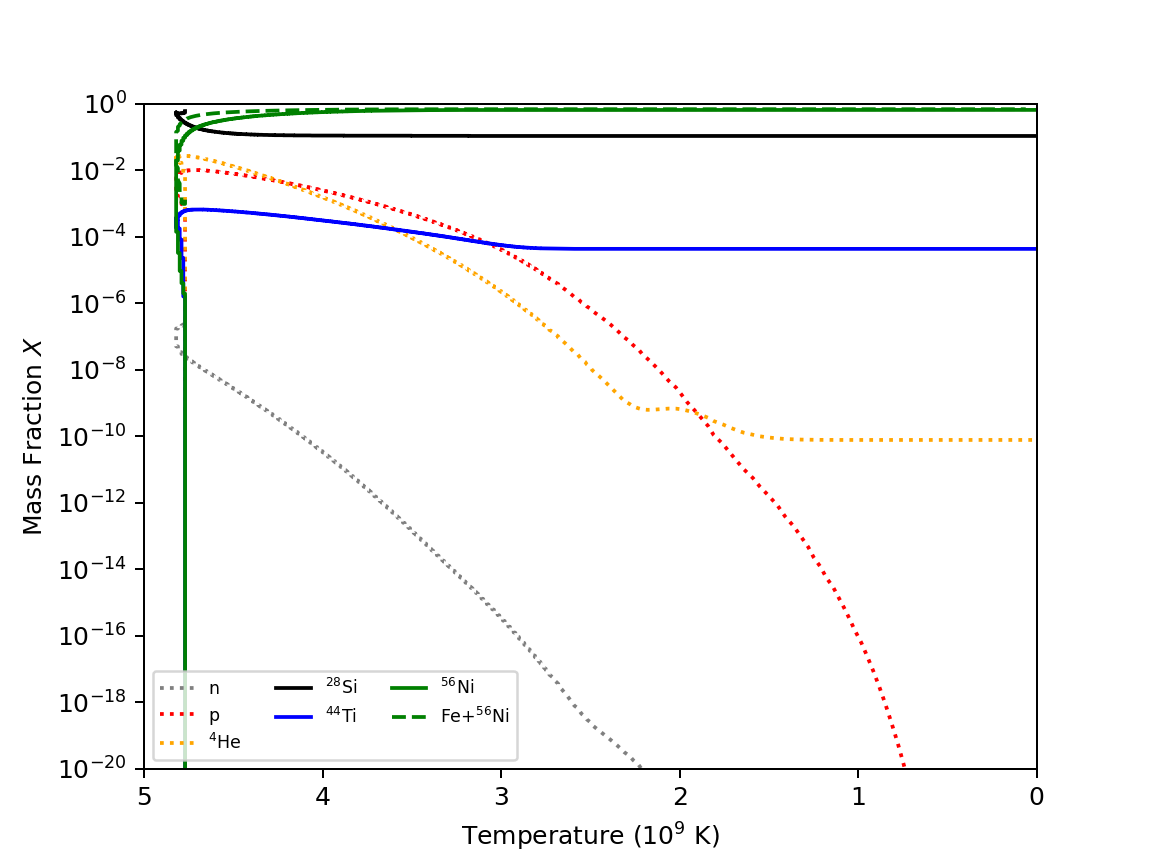}{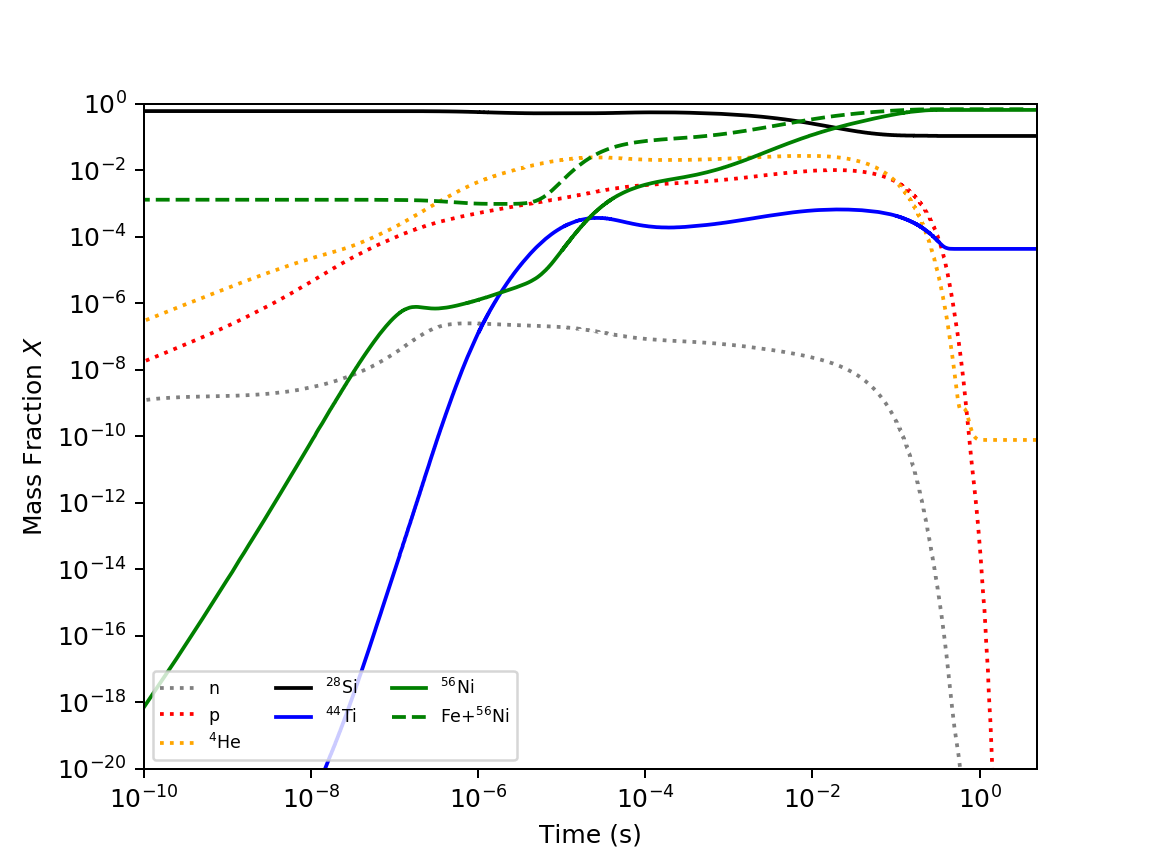}
\caption{Same as Figure~\ref{plaw_reg1}, but using best-fit power law trajectories of the  region 2 particles.}
\label{plaw_reg2}
\end{figure*}

\begin{figure*}
\epsscale{1.14}
\plottwo{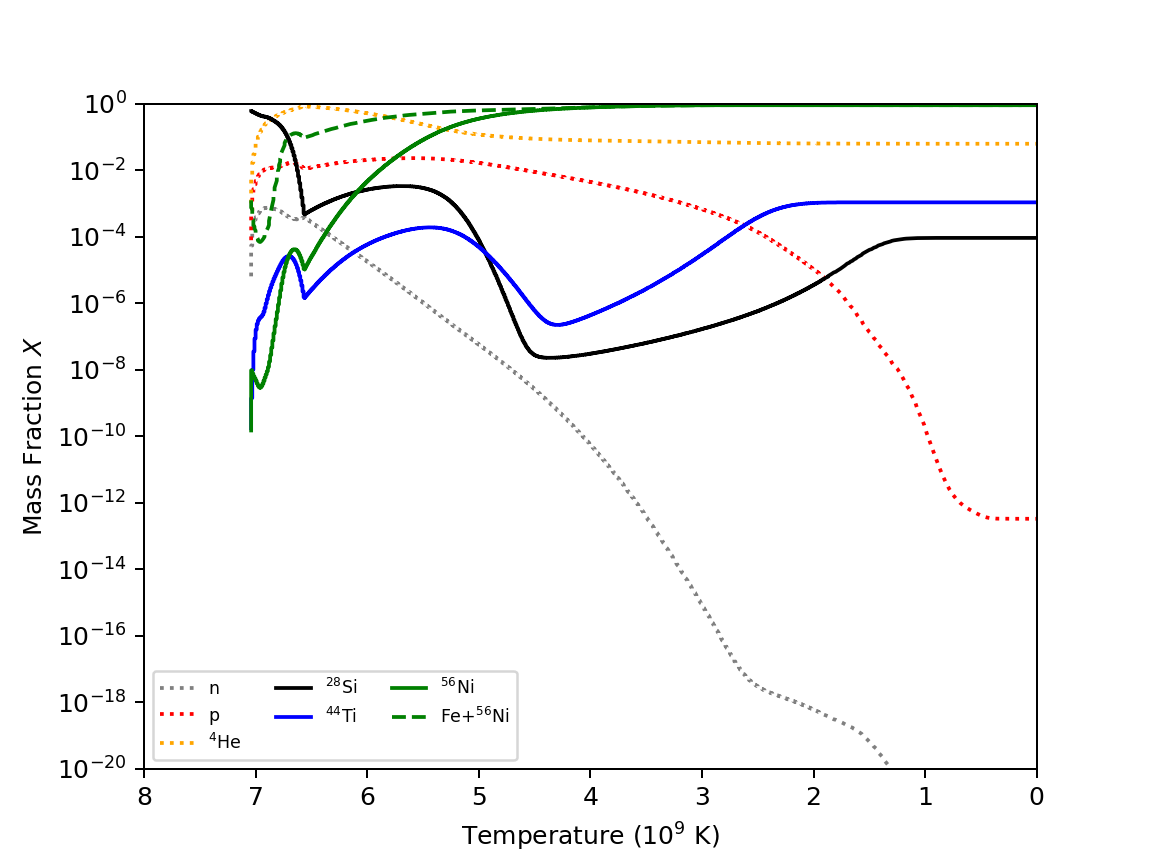}{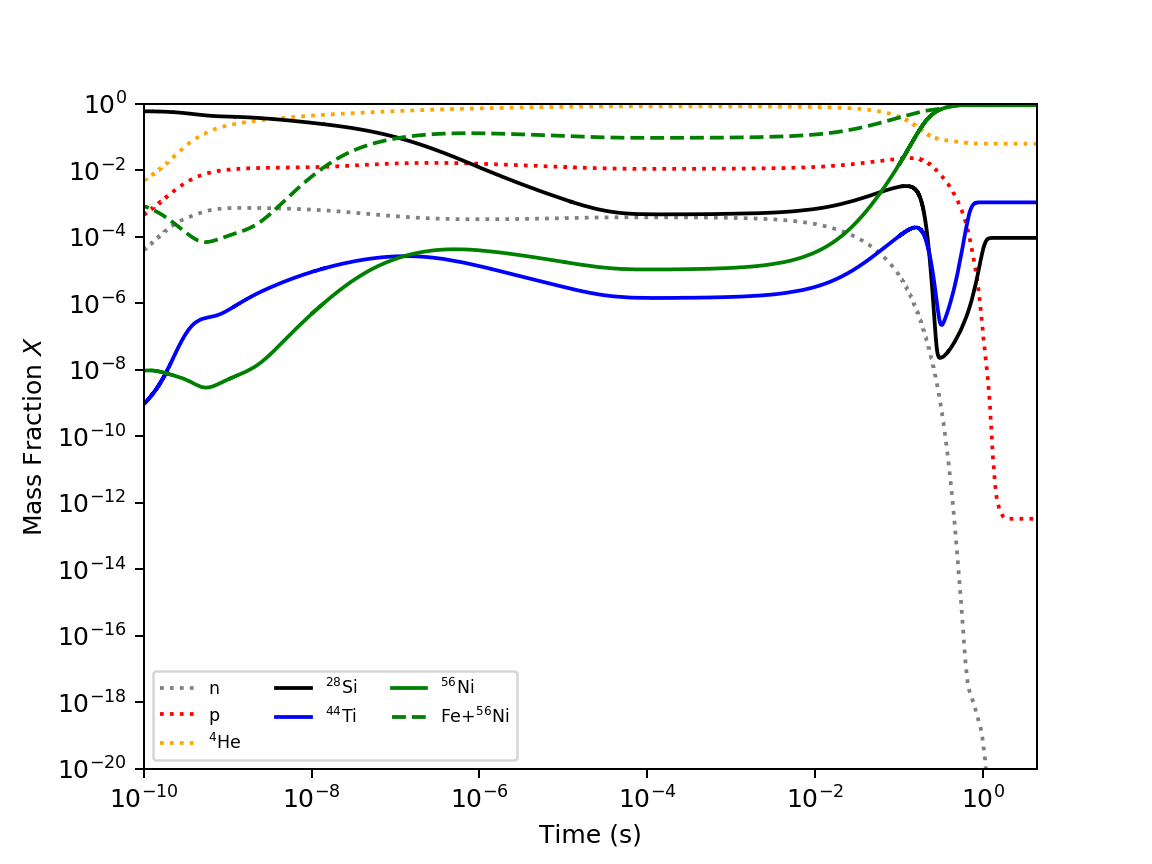}
\caption{Same as Figure~\ref{plaw_reg1}, but using best-fit power law trajectories of the region 3 particles.}
\label{plaw_reg3}
\end{figure*}

\begin{figure*}
\epsscale{1.14}
\plottwo{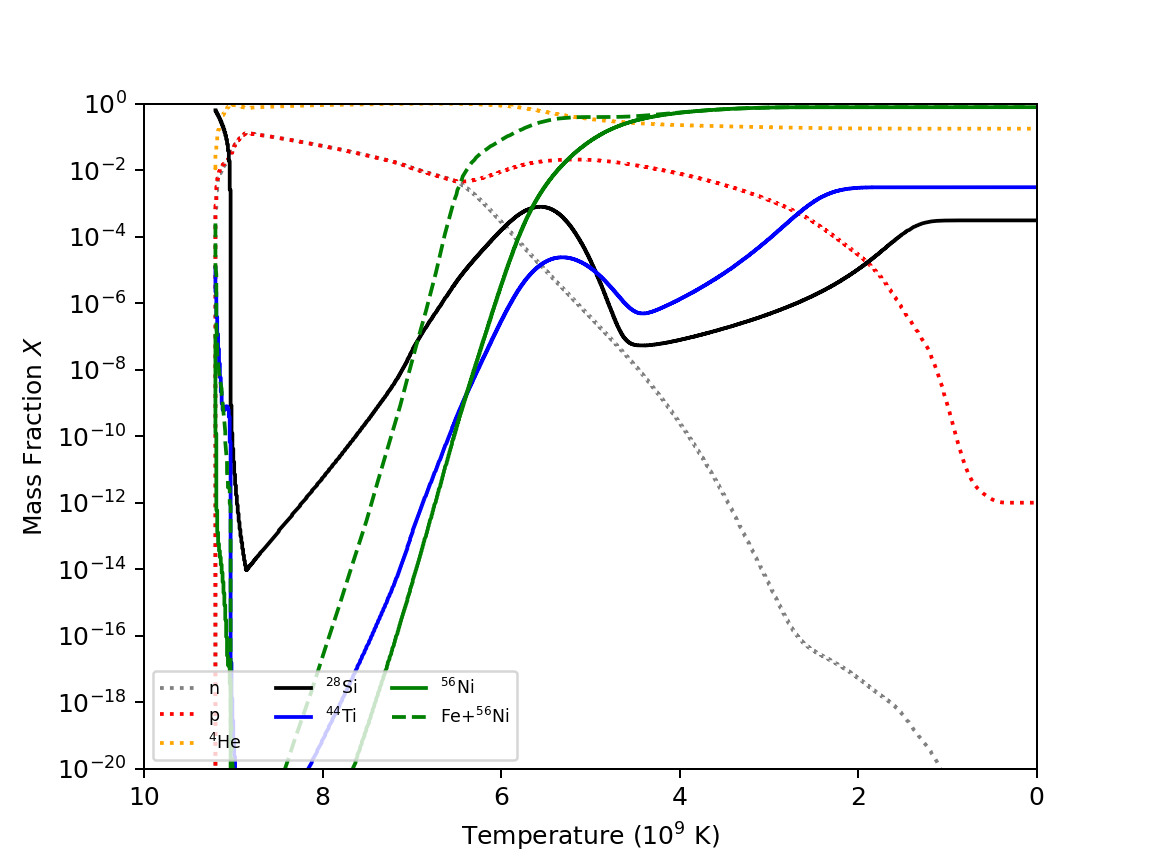}{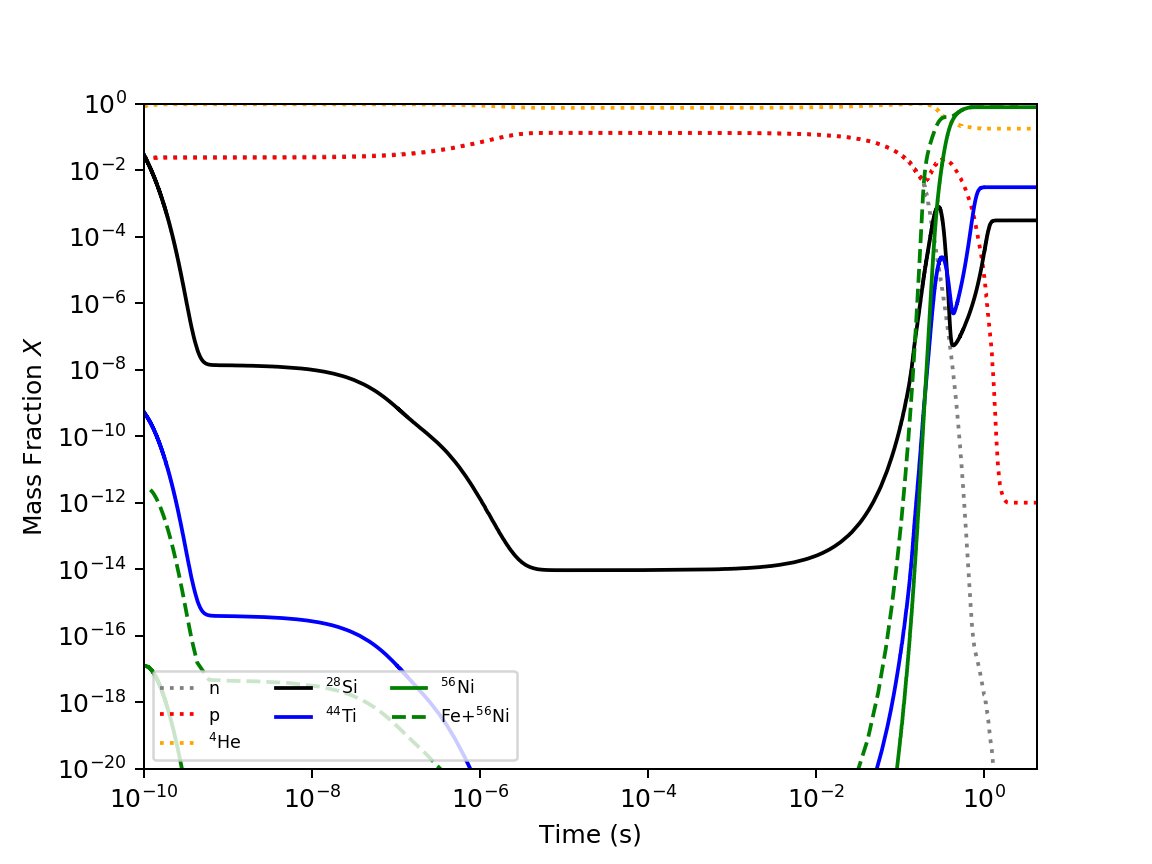}
\caption{Same as Figure~\ref{plaw_reg1}, but using best-fit power law trajectories of the region 4 particles.}
\label{plaw_reg4}
\end{figure*}

To understand the dependencies of the yields on the temperature and density evolution better, we performed nucleosynthesis calculations for trajectories corresponding to the mean trajectories in the four regions.  For each of these trajectories, we start with initial mass fractions of $X(\iso{O}{16}) = 0.7$, $X(\iso{Si}{28}) = 0.2997$, and $X(\iso{Fe}{56}) = \sci{1.3}{-3}$.  For each region, we use two trajectories:  the first based on the mean temperature/density evolution from our simulations and the second based on the best-fit power laws to each simulated trajectory. Exponential trajectories were found to be very poor fits to the evolution. The final yields of each of these models can be found in table~\ref{abuns_table}.  The final yields can differ between our simulation trajectories and the simple power laws by more than a factor of 2. In the case of \iso{Ti}{44}, production varies by an order of magnitude in region 4. For region 1, these differences demonstrate just how sensitive the yields are in the incomplete-burning, Si-rich phase.  In regions 3 and 4, the differences are caused because the pileup alters the temperature considerably when nuclear burning is still active.  Figures~\ref{traj_reg1}--\ref{traj_reg4} show the evolution (based on the simulation trajectories) of the \iso{Ti}{44}, \iso{Ni}{56}, \iso{Si}{28}, neutron, proton, and $\alpha$-particle abundances both plotted against time and plotted against temperature.  Figures~\ref{plaw_reg1}--\ref{plaw_reg4} show this same evolution for the best-fit power law models.

\begin{deluxetable*}{l|cc|cc|cc|cc}
\tablewidth{0pt}
\tablecaption{Comparison of selected final abundances for nucleosynthesis calculations using the mean trajectories from each region and the best-fit power law trajectories. \label{abuns_table}}
\tablehead{\colhead{} & \colhead{Region 1} & \colhead{Region 1} & \colhead{Region 2} & \colhead{Region 2} & \colhead{Region 3} & \colhead{Region 3} & \colhead{Region 4} & \colhead{Region 4} \\
\colhead{} & \colhead{Power Law} & \colhead{Mean Traj.} & \colhead{Power Law} & \colhead{Mean Traj.} & \colhead{Power Law} & \colhead{Mean Traj.} & \colhead{Power Law} & \colhead{Mean Traj.}}
\startdata
n & \sci{2.16}{-49} & \sci{1.14}{-48} & \sci{3.66}{-57} & \sci{1.54}{-55} & \sci{1.41}{-44} & \sci{9.52}{-45} & \sci{2.56}{-42} & \sci{3.79}{-43} \\
p & \sci{3.46}{-31} & \sci{6.11}{-34} & \sci{1.19}{-26} & \sci{3.04}{-29} & \sci{3.32}{-13} & \sci{2.58}{-15} & \sci{1.01}{-12} & \sci{2.40}{-14} \\
\iso{He}{4} & \sci{2.25}{-16} & \sci{1.20}{-16} & \sci{7.78}{-11} & \sci{7.61}{-11} & \sci{6.24}{-2} & \sci{1.09}{-1} & \sci{1.77}{-1} & \sci{2.80}{-1} \\
\iso{Si}{28} & \sci{5.38}{-1} & \sci{7.22}{-1} & \sci{1.07}{-1} & \sci{2.37}{-1} & \sci{9.22}{-5} & \sci{1.41}{-4} & \sci{3.10}{-4} & \sci{2.09}{-4} \\
\iso{Ti}{44} & \sci{6.54}{-6} & \sci{1.91}{-7} & \sci{4.31}{-5} & \sci{5.22}{-5} & \sci{1.08}{-3} & \sci{2.14}{-3} & \sci{3.08}{-3} & \sci{4.93}{-2} \\
\iso{Ni}{56} & \sci{8.46}{-6} & \sci{1.56}{-6} & \sci{6.50}{-1} & \sci{3.93}{-1} & \sci{9.00}{-1} & \sci{8.42}{-1} & \sci{7.86}{-1} & \sci{5.75}{-1} \\
Fe+\iso{Ni}{56} & \sci{1.13}{-3} & \sci{1.15}{-3} & \sci{6.82}{-1} & \sci{4.18}{-1} & \sci{9.01}{-1} & \sci{8.44}{-1} & \sci{7.89}{-1} & \sci{5.88}{-1} \\
\enddata
\end{deluxetable*}

Region 1 has the lowest peak temperature, and the pileup occurs when the ejecta is already quite cool.  In this case, the bump in temperature has very little effect on the final yields.  However, the initial rapid temperature evolution (somewhere in between an exponential decay and a power law decay) means that the burning time frame is much more abrupt in our simulations with respect to a power law fit, producing less \iso{Ti}{44}.  This causes our region 1 yields to be lower than those expected from a power law profile. The same abrupt evolution occurs in region 2, causing slightly different final yields.  As with region 1, the increase in the temperature when the material is hit by the reverse shock has only a minimal effect on the final yields.

Regions 3 and 4 are hotter than the first two regions and most of the \iso{Ti}{44} production occurs when the trajectory is well mimicked by a power law.  However, the reverse shock drives the temperatures in these two regions sufficiently high to alter the late-time burning, altering the final yields.  In these cases, this phase increases the net \iso{Ti}{44} production.  

In the very broad $\alpha$-rich freezeout region both \iso{Ni}{56} and \iso{Ti}{44} have a weak dependence on thermodynamic conditions and converge on a low \feti{} ratio, the exact value of which can be altered by dynamics that deviate from an analytic trajectory.  The transition to the QSE-leakage chasm is extremely sharp in temperature. Within the chasm, many particles have $\log(\feti) \sim 3.9$, and very little material with intermediate \feti{} is produced. On the Si-burning side of the chasm \iso{Ti}{44} abruptly rises, though not to the level seen in the $\alpha$-rich freezeout. At lower temperatures, production of both \iso{Ni}{56} and \iso{Ti}{44} are similar and much reduced, but the presence of Fe in the star's original composition conspires to maintain a ratio of $\log(\feti) \sim 2.6$. This low temperature, low \iso{Ni}{56} material should, in principle, be distinguishable from high temperature production with a similar $\log(\feti) \sim 2.6$ by the presence of significant amounts of Si. For material in a Population~I supernova with $\rho (T_\mathrm{peak}) < \sci{1.0}{8}\ \mathrm{g\ cm^{-3}}$, a broadly bimodal \feti{} distribution like that observed in Cas~A is a natural consequence of the nucleosynthesis. Since the formation of this bimodal distribution in \feti{} is so robust to details of the trajectories, we should find observationally that this distribution is a common feature of supernova remnants.

 
These studies of the yields production confirm the basic trends in \cite{Magkotsios10} with the yields depending on different physics in different regions, but the final results \emph{do} depend on the temperature/density evolution of the ejecta.  In some cases, an exponential decay followed by a power law decay may be a better fit for the trajectories, but neither of these trajectories account for the pileup and heating of the ejecta which, for some cases, can significantly alter the yields.  It is important to gain intuition from simple trajectories, but quantitative results will require studies using simulated temperature/density evolution.




\section{Summary}
\label{sec:summary}

\begin{figure}
\epsscale{1.30}
\plotone{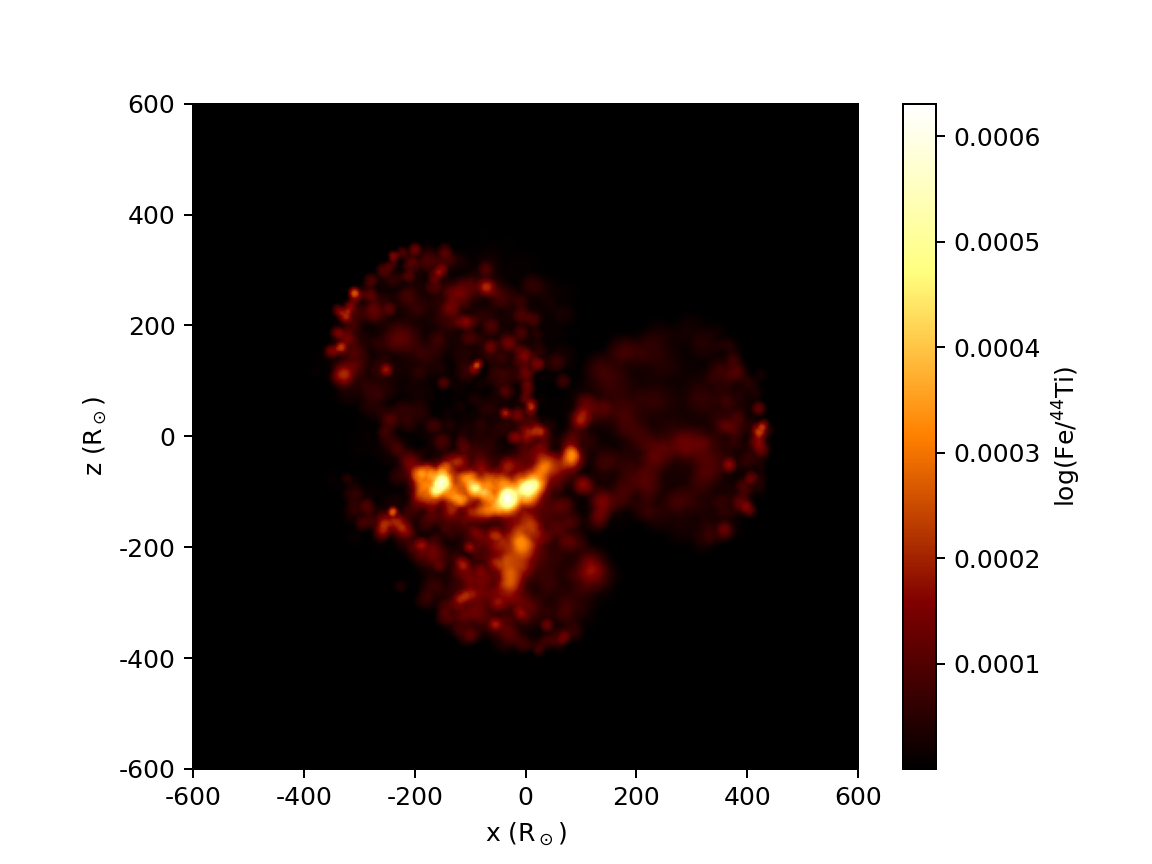}
\caption{Rendering in 3D of the \iso{Ti}{44} distribution tracing the asymmetries in the supernova engine.  The color intensity indicates the log of the mass fraction of \iso{Ti}{44}.  The distribution of the \iso{Ni}{56} traces the same regions.}
\label{fig:tirender}
\end{figure}
\begin{figure}
\epsscale{1.30}
\plotone{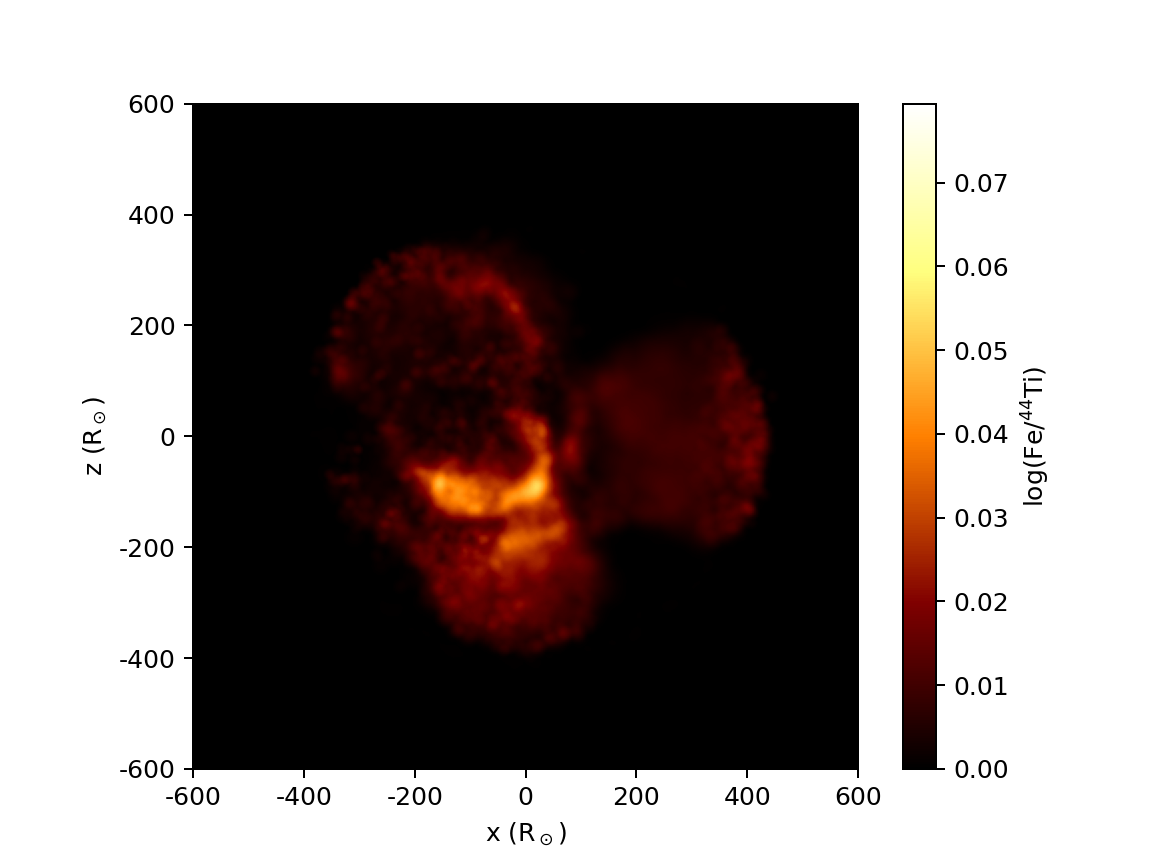}
\caption{Rendering in 3D of the total iron distribution, consisting of all Fe isotopes plus \iso{Ni}{56}.  As with figure~\ref{fig:tirender}, the color intensity indicates the log of the mass fraction.  The total iron production in our model extends beyond the \iso{Ti}{44} distribution (see figure~\ref{fig:tirender}).}
\label{fig:ironrender}
\end{figure}
\begin{figure}
\epsscale{1.30}
\plotone{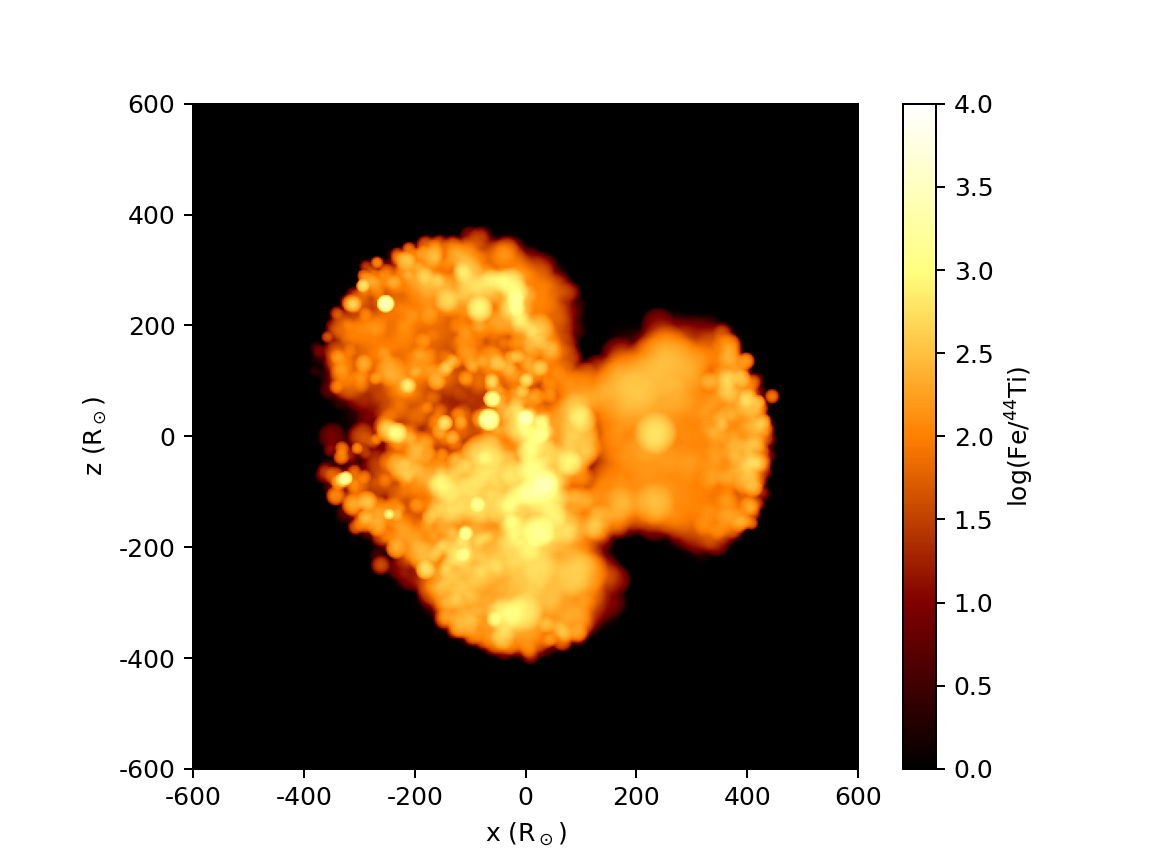}
\caption{Rendering in 3D of the spatial \feti{} ratio in our model.  The color intensity indicates the log of the ratio.}
\label{fig:fetirender}
\end{figure}

In this paper, we studied the \iso{Ti}{44} and \iso{Ni}{56} production in a three-dimensional supernova explosion model.  For the most part, these two isotopes are both produced in the innermost ejecta of the explosion, and their distributions reflect the asymmetries in the supernova engine.  Figure~\ref{fig:tirender} shows the \iso{Ti}{44} distribution at late times in our model.  The asymmetries in this distribution trace asymmetries in the explosive engine.  The total iron distribution (shown in figure~\ref{fig:ironrender}) extends beyond the \iso{Ti}{44}, but also traces the asymmetries in the explosive engine. Figure~\ref{fig:fetirender} shows the \feti{} ratio in regions with \iso{Ti}{44} mass fraction $X(\iso{Ti}{44}) > 10^{-6}$. As shown in Figures~\ref{totalmass1dhist} and \ref{ironmass1dhist}, the \feti{} ratios of particles in the explosion form a largely bimodal distribution with peaks at $\log(\feti) \sim 2.5$ and $\log(\feti) \sim 3.9$.  This formation of such a distribution is robust under a range of conditions, so it would be expected as a common feature of supernova remnants.

The \feti{} ratio provides additional clues into the nature of the supernova explosion.  We have shown how the density and temperature evolution affects this ratio.  Asymmetric supernovae like our model produce a wide range of conditions that can produce \feti{} ratios that vary from as low as $\log(\feti) \sim 1.0$ (up to $\sim 10\%$ of the total mass of iron in \iso{Ti}{44}) to high $\log(\feti)$ values exceeding 5.0 (very little \iso{Ti}{44}).  In general, the trends in these yields follow the trends studied by \cite{Magkotsios10} where the $\alpha$-rich freezeout region produces the most \iso{Ti}{44} (lowest \feti{} value).  Material produced in the chasm or in the incomplete-burning (i.e., silicon-rich) region produces higher \feti{} values.  High \iso{Ti}{44} abundances in maps of supernova remnants like those done by NuSTAR \citep{Magkotsios10} probe strong shock regions.

However, the density and temperature evolution can vary dramatically from the simple power law or exponential decays studied in most parameter studies.  In our asymmetric explosions, shocks can cause some ejecta to re-heat while still undergoing nuclear burning, and these alterations in the evolution can significantly alter the final yields.  Detailed three-dimensional studies such as those presented here are essential in using details in the observed abundances to probe the exact conditions of the shocks.

In this work, we also demonstrate that full 3D supernova models are capable of developing stochastic large-scale asymmetries driven by convection from fallback of material onto the proto-neutron star. Since this asymmetry formed on its own and was not imposed by the model setup, it seems likely that other 3D models could independently develop similar behaviors.  Indeed, this behavior could be an important universal feature of supernovae.  The uniformity of this nucleosynthetic behavior under the effect of stochastic motions from convective processes implies that the details of circulation of material from any early time process, whether convection from the engine, changing gravitational potential, or pre-explosion progenitor convection, are not the dominant controllers of \feti{}. We should expect a bimodal \feti{} distribution in the majority of supernovae with ejecta that reaches NSE temperatures. Uniformly low \feti{} would indicate that only very high density material reached NSE temperatures. In order for lower density material to produce only low \feti{}, little QSE processing can take place despite material being heated to NSE. Therefore a low \feti{} unimodal distribution requires a very massive progenitor to produce the requisite densities and a strongly asymmetric explosion to excavate the high density material without heating shallower, lower density regions to NSE temperatures.

\begin{figure}
\epsscale{1.15}
\plotone{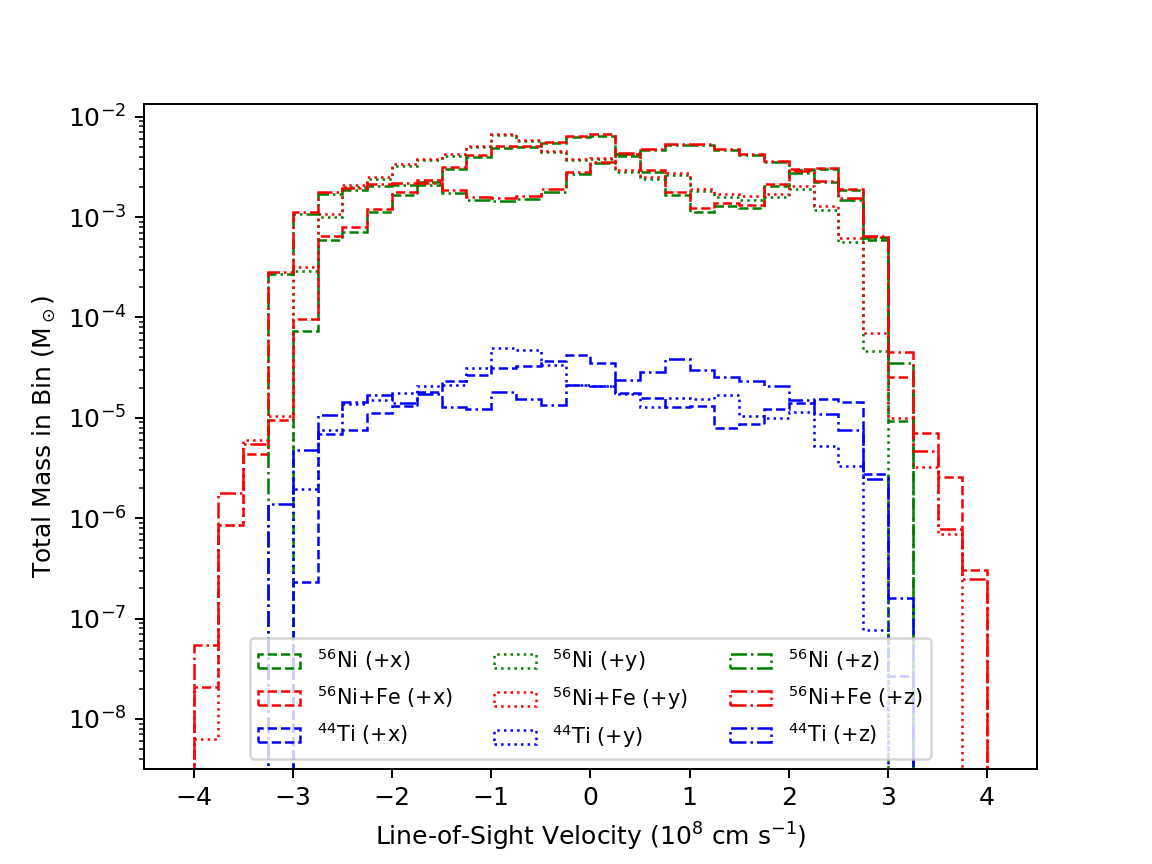}
\caption{A visualization of \iso{Ni}{56}, \iso{Ti}{44}, and total iron distribution plotted against line-of-sight velocity for a sample of three lines of sight (parallel to the $x$ axis, parallel to the $y$ axis, and parallel to the $z$ axis.}
\label{fig:velocity}
\end{figure}

Many remnants are too distant to be resolved with high-energy telescopes.  However, we can use the Doppler broadening of lines to study the structure in the ejecta.  Figure~\ref{fig:velocity} shows the velocity distribution of the \iso{Ti}{44}, \iso{Ni}{56}, and the $\mathrm{Fe} + \iso{Ni}{56}$ ejecta for a few different lines of sight in our simulation.  The structure in these velocity distributions can provide evidence of asymmetries in the explosive engine.  Hard X-ray and $\gamma$-ray missions that have good energy resolution may be able to resolve these features.  As the sample of such observations grow, we will be able to study the level of asymmetry in core-collapse supernovae.

\begin{figure}
\epsscale{1.10}
\plotone{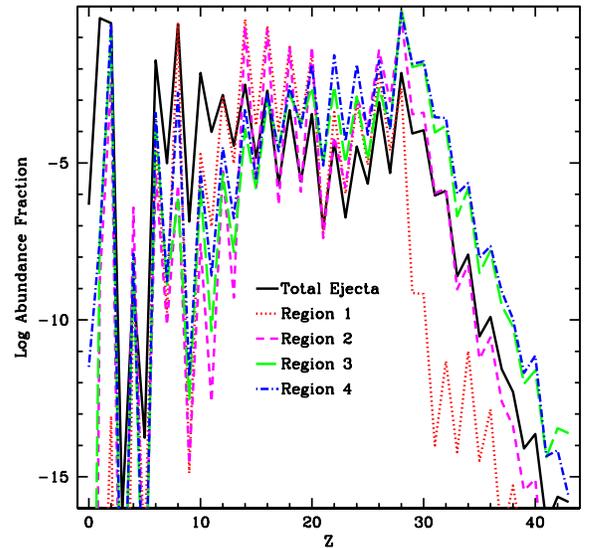}
\caption{Total and specific abundances plotted against isotope proton number $Z$ for all four discussed regions in our explosive ejecta.}
\label{fig:fullabun}
\end{figure}

Other abundances also depend on the shock strength (figure~\ref{fig:fullabun}), and observing the variations of these abundances in different regions of the remnant can also probe properties of the explosion.  Figure~\ref{fig:fullabun} shows both the total yield distribution as well as the yield distributions in each of our four regions independently. The hotter regions visibly produce heavier elements in greater abundance, but we defer a detailed study of these abundances to a later paper. Our models do not include late-time engine interactions that are seen in many multi-dimensional models \citep[e.g.,][]{2017ApJ...843....2H}.  These effects will also alter the yields, and much more work is needed to understand the full yields from supernovae.

\acknowledgments
Support for this work was provided by the National Science Foundation award 1615575. This work was supported, in part, by the US Department of Energy through the Los Alamos National Laboratory. Los Alamos National Laboratory is operated by Triad National Security, LLC, for the National Nuclear Security Administration of the U.S. Department of Energy (Contract No.~89233218CNA000001). Research presented in this article was supported by the Laboratory Directed Research and Development program of Los Alamos National Laboratory under project number 20190021DR.

\software{NumPy \citep{numpy1,numpy2}, SciPy \citep{scipy}, Matplotlib \citep{matplotlib}, SPLASH \citep{Price2007}, Py-SPHViewer \citep{pysphviewer}}

\bibliographystyle{aasjournal}
\bibliography{references}

\end{document}